\documentclass[iop]{emulateapj}

\usepackage{graphicx}
\usepackage{color}
\usepackage{amsmath,amssymb}
\usepackage{verbatim}
\usepackage{enumerate}
\usepackage{multirow}
\usepackage{lineno}
\usepackage[normalem]{ulem}
\usepackage{hyperref}

\newcommand{\SNIa}{SN\,Ia}
\newcommand{\SNeIa}{SNe\,Ia}

\newcommand{\PSNID}{{\tt PSNID}}
\newcommand{\urlDR}{{\tt https://des.ncsa.illinois.edu/releases/sn}}

\newcommand{\Diff}{{\tt DiffImg}}
\newcommand{\SNANA}{{\tt SNANA}}
\newcommand{\SMP}{{\tt SMP}}

\begin{document}

{
\vspace*{-\headsep}\vspace*{\headheight}
\footnotesize \hfill FERMILAB-PUB-18-632-AE\\
\vspace*{-\headsep}\vspace*{\headheight}
\footnotesize \hfill DES-2016-0157
}

\title{First Cosmology Results Using Type Ia Supernovae From the Dark Energy Survey: Survey Overview and Supernova Spectroscopy}
\shortauthors{D'Andrea et al.}

\def\andname{}

\author{
C.~B.~D'Andrea\altaffilmark{1},
M.~Smith\altaffilmark{2},
M.~Sullivan\altaffilmark{2},
R.~C.~Nichol\altaffilmark{3},
R.~C.~Thomas\altaffilmark{4},
A.~G.~Kim\altaffilmark{4},
A.~M\"oller\altaffilmark{5,6},
M.~Sako\altaffilmark{1},
F.~J.~Castander\altaffilmark{7,8},
A.~V.~Filippenko\altaffilmark{9,10}
% ...
\\ \vspace{0.2cm} (DES Collaboration) \\
}

\affil{$^{1}$ Department of Physics and Astronomy, University of Pennsylvania, Philadelphia, PA 19104, USA}
\affil{$^{2}$ School of Physics and Astronomy, University of Southampton,  Southampton, SO17 1BJ, UK}
\affil{$^{3}$ Institute of Cosmology and Gravitation, University of Portsmouth, Portsmouth, PO1 3FX, UK}
\affil{$^{4}$ Lawrence Berkeley National Laboratory, 1 Cyclotron Road, Berkeley, CA 94720, USA}
\affil{$^{5}$ ARC Centre of Excellence for All-sky Astrophysics (CAASTRO)}
\affil{$^{6}$ The Research School of Astronomy and Astrophysics, Australian National University, ACT 2601, Australia}
\affil{$^{7}$ Institut d'Estudis Espacials de Catalunya (IEEC), 08034 Barcelona, Spain}
\affil{$^{8}$ Institute of Space Sciences (ICE, CSIC),  Campus UAB, Carrer de Can Magrans, s/n,  08193 Barcelona, Spain}
\affil{$^{9}$ Department of Astronomy, University of California, Berkeley, CA 94720-3411, USA}
\affil{$^{10}$ Miller Senior Fellow, Miller Institute for Basic Research in Science, University of California, Berkeley, CA  94720, USA}
% ...

\begin{abstract}
We present spectroscopy from the first three seasons of the Dark Energy Survey Supernova Program (DES-SN).  We describe the supernova spectroscopic program in full: strategy, observations, data reduction, and classification.  We have spectroscopically confirmed 307 supernovae, including 251 type Ia supernovae (\SNeIa) over a redshift range of $0.017 < z < 0.85$.  We determine the effective spectroscopic selection function for our sample, and use it to investigate the redshift-dependent bias on the distance moduli of \SNeIa\ we have classified.  We also provide a full overview of the strategy, observations, and data products of DES-SN, which has discovered 12,015 likely supernovae during these first three seasons.  The data presented here are used for the first cosmology analysis by DES-SN (`DES-SN3YR'), the results of which are given in \citet{DES-SN2018}.

\end{abstract}

\section{INTRODUCTION}
\label{sec:intro}

Type Ia supernovae (\SNeIa) have fundamentally changed our understanding of the universe.  It is through their utility as accurate distance indicators that the High-Z Supernova Search Team \citep{Riess1998} and the Supernova Cosmology Project \citep{Perlmutter1999} were able to make the groundbreaking discovery that the expansion of the universe is accelerating.  To date, the nature of the substance causing this phenomenon, commonly referred to as `dark energy,' remains unknown.   

The quest for understanding the cause of the acceleration and constraining the models that describe it have motivated ever-improving supernova searches over the past two decades.  At redshift $z\leq 1$ these cosmology-oriented programs include the Supernova Legacy Survey \citep[SNLS;][]{Astier2006, Sullivan2011, Conley2011}, the Sloan Digital Sky Survey-II Supernova Program \citep[SDSS-II;][]{Frieman2008, Kessler2009a, Sako2008, Sako2018}, ESSENCE \citep{Wood-Vasey2007,Miknaitis2007,Narayan2016}, and more recently, Pan-STARRS \citep{Scolnic2014, Rest2014, Scolnic2018}.  The low-redshift sample necessary for anchoring the Hubble diagram includes Cal\'an-Tololo \citep{Hamuy1996}, several CFA samples \citep{Riess1999,Jha2006,Hicken2009,Hicken2012}, the Carnegie Supernova Project \citep[CSP;][]{Contreras2010}, and, more recently, the homogeneous Foundation Survey \citep{Foley2018}.  Nearly all observations of \SNeIa\ at $z>1.1$ are obtained from space, with only a few dozen well-observed objects to date \citep{Riess2004,Riess2007,Suzuki2012,Riess2018}.

The surveys described above all obtain distance measurements from light-curve fits to cadenced multi-color photometry \citep{Phillips1993,Riess1996,Tripp1998}.  But to define the sample of \SNeIa\ used in a cosmological analysis requires a parallel spectroscopic follow-up program.  This allows the survey to differentiate observed transients between \SNeIa\ and other classes of supernova, while also obtaining precise redshifts for the objects.  Thus, the spectroscopic program determines what data are included in a Hubble diagram, as well as the position of these data along one axis.

For supernova surveys in the range of $0.1<z<1.0$, the spectroscopic program typically requires more observing time, and on larger telescopes, than does the entire photometric observing program---all of this to classify a small subset of detected transients.  For example, SNLS used $\sim$~900 hours of spectroscopy on 8-10m class telescopes to spectroscopically classify 285~\SNeIa\ in the first three years of their survey \citep{Howell2005,Bronder2008,Balland2009,Ellis2008}, compared to 779 hours of good-quality photometry \citep{Guy2010}.  Similarly large resources were dedicated by other large programs in this redshift range, such as ESSENCE \citep[213 \SNeIa;][]{Matheson2005,Foley2009,Narayan2016}, Pan-STARRS1 \citep[361 \SNeIa;][]{Rest2014, Scolnic2018} and SDSS-II \citep[500 \SNeIa;][]{Zheng2008, Sako2018}.

In this paper we describe the supernova spectroscopy program for the first three seasons of the Dark Energy Survey - Supernova Program (DES-SN), and give an overview of the survey and its operations.  This is part of a series of companion papers supporting the first cosmological analysis of spectroscopically-classified \SNeIa\ from DES-SN (`DES-SN3YR').  These include detailed papers on aspects of SN search and discovery \citep{Kessler2015, Goldstein2015, Morganson2018}, our photometry pipeline \citep{Brout2018-PHOT}, photometric calibration \citep{Burke2018,Lasker2018}, simulations \citep{Kessler2018}, and a technique to account for simulation bias \citep{KS2017}.  Our analysis methodology and systematic uncertainties are presented by \citet{Brout2018-SYS}, and these results are used to constrain cosmology \citep{DES-SN2018} and the Hubble constant \citep{Macauley2018}.  The DES-SN3YR constraints are combined with other DES probes in \citet{DESALL2018}.  A new Bayesian Hierarchical Model for supernova cosmology is tested by \citet{Hinton2018}.

The format of the paper is as follows.  In Section~\ref{sec:DES} we describe the strategy and status of the overall DES-SN observing program.  In Section~\ref{sec:PhotCands} we describe how SN candidates are defined and extracted from the data, while Section~\ref{sec:Spec} details the spectroscopic follow-up campaign for each observatory used in our program.  In Section~\ref{sec:ssf} we derive the effective spectroscopic selection function from the classifications obtained by our program, essential for understanding the biases in a spectroscopically derived SN Hubble diagram from DES.  We conclude in Section~\ref{sec:conc} by looking toward future releases and analyses of DES data.

\section{The DES Supernova Program}
\label{sec:DES}

\subsection{Dark Energy Survey}
The Dark Energy Survey \citep{DES2016} is a 6 year, $\sim570$ night survey using the 4-meter Blanco telescope at Cerro-Tololo Inter-American Observatory (CTIO) in Chile.  It uses the Dark Energy Camera \citep[DECam;][]{Flaugher2015}, a 520 megapixel wide-field imager with a $2.2\arcdeg$ field of view and deep-depleted CCDs, giving it excellent quantum efficiency out to 1 micron.  Commissioning of the camera began in September 2012, and a period of data-taking by DES called Science Verification (`SV') was carried out from November 2012 through February 2013.  The first season of the survey (`Y1') began on August 31, 2013, and the third season (`Y3') ended February 12, 2016.

DES is designed as a Stage 3 dark-energy experiment according to the Dark Energy Task Force (DETF) Figure of Merit \citep{Albrecht2006}, increasing the constraints on the $w_0-w_a$ plane by a factor of a few.  It combines four probes of dark energy -- weak lensing, large-scale structure, galaxy clusters, and \SNeIa\ -- into one experiment sharing a common instrument, allowing for consistent calibration, validation, and a better understanding of systematic errors in the combined analysis.  DES is split into two distinct observing modes:  the wide-area survey (DES-wide), observing 5000 square degrees in $grizY$ to a 5$\sigma$ depth of $\sim 23.5$; and DES-SN.   

The observing strategy for DES-SN is optimized for the purposes of \SNIa\ cosmology.  With this in mind, different observing strategies were explored by \citet{Bernstein2012}.  The selected and implemented strategy for DES-SN is a 10-field hybrid-depth survey, designed to obtain a few thousand well-observed light curves of \SNeIa\ over a redshift range $0.2<z<1.2$.  

There are three defining aspects of DES-SN.  The first is the excellent $z$-band response of DECam owing to the deep-depleted CCDs \citep{Diehl2014}.  This allows for rest-frame optical light curves of $z\approx1$~\SNeIa\ to be well-measured.  The second is excellent calibration, as this is the largest systematic uncertainty in \SNIa\ cosmology \citep{Scolnic2018}.  The DES Science Requirements state that the survey must be calibrated to 0.5\% in its absolute calibrations and colors.

The third defining aspect is photometric classification of \SNeIa.  The field-of-view of DECam is much larger than that of any previous camera on a similarly sized telescope, allowing DES-SN to observe an unprecedented area for its depth.  Thus, given the quantity of faint SNe that DES-SN discovers, any realistic spectroscopic resource allocation will only permit spectroscopic classification for a small fraction of these SNe.  To make optimal use of the DES-SN data we therefore will rely on photometric classification for our primary cosmology analysis.  This does not remove the need for spectroscopic follow-up observations of live SNe, but rather places different priorities on the follow-up program, as explained in Section~\ref{sec:Spec}.

In the remainder of this section we first describe in detail the first three seasons of the DES-SN observing program.

\subsection{Exposure Time and Depth}
\label{sec:exptime}

For a fixed amount of observing time there is a direct trade-off between depth and area.  Based on simulations of different survey strategies
\citep{Bernstein2012}, DES-SN has been designed to have fields of two different depths:  eight `shallow' and two `deep' fields, where each field is a single pointing of DECam.  The deep fields serve to extend the redshift range of cosmologically useful \SNeIa\ out to $z\approx1.2$, while the more numerous shallow fields add volume and numbers at intermediate redshifts.  The total exposure time for each filter and the median limiting magnitude for both the deep and shallow fields are given in Table~\ref{tbl:tobs}.  Unlike DES-wide, DES-SN observes in only $griz$, as $Y$-band exposures provide too little additional information to justify the significant added cost in exposure time.  Longer observations are split into a number of shorter exposures and coadded (e.g., 11 exposures for the 1hr per-epoch $z$-band deep fields).  We note that the limiting magnitude is derived from artificially inserting supernovae into our processing pipeline and determining the magnitude at which 50\% of all such objects are recovered \citep{Kessler2015}.  As such, these limits take into account real observing conditions and template noise, and are $0.5$ to $1$ mag shallower than when computed with the DECam Exposure Time Calculator (ETC)\footnote{\url{http://www.ctio.noao.edu/noao/node/5826}}, as was also discussed by \citet{Forster2016}.  Throughout the text, we refer to an observation of one field, in one filter, on one night, as a `filter-epoch'.

\begin{deluxetable} {c*{7}{c}}
\tablewidth{0pt}
\tabletypesize{\scriptsize}
\tablecaption{Exposure Times\label{tbl:tobs}}
\tablehead{
  \multirow{2}{*}{Filter} &
  \multicolumn{3}{c}{Shallow Field}	& &
  \multicolumn{3}{c}{Deep Field} \\
  \cline{2-4} \cline{6-8} 
  & $t_{\textrm{exp}}\tablenotemark{a}$ & $N_{\textrm{exp}}$\tablenotemark{b} & Depth\tablenotemark{c}  &
  & $t_{\textrm{exp}}$\tablenotemark{a} & $N_{\textrm{exp}}$\tablenotemark{b} & Depth\tablenotemark{c}
}
\startdata
$g$        & 175 & 1 & 23.7  & & 600  & 3  & 24.6 \\
$r$        & 150 & 1 & 23.6  & & 1200 & 3  & 24.8 \\
$i$        & 200 & 1 & 23.5  & & 1800 & 5  & 24.7 \\
$z$        & 400 & 2 & 23.3  & & 3630 & 11 & 24.4
\enddata
\tablenotetext{a}{Total exposure time per filter-epoch (in seconds).}
\tablenotetext{b}{Number of exposures per filter-epoch.} 
\tablenotetext{c}{Median limiting magnitude per pointing over the first 3 seasons of DES-SN, defined as the magnitude at which 50\% of fake supernovae inserted into pipeline are recovered by difference imaging \citep{Kessler2015}.}
\end{deluxetable}

DECam has a 2.2$^{\circ}$ diameter field-of-view and an observable area (excluding chip gaps) of 2.7 deg$^2$.  This gives the DES-SN program a total observing area of 27 deg$^2$, nearly 7 times the area of SNLS.  DES-SN does not dither over the gaps, since filling these in decreases the area repeatedly observed on the subsequent epoch.  Dithers of order of a few arcseconds are carried out, allowing instrumental artifacts to be corrected in processing.  Since dithers do not cross the chip gap, any object in a field appears on only one chip; therefore, our processing pipeline treats each chip independently.

\subsection{Field Locations}
\label{sec:fieldpos}

The ten DES-SN fields are grouped in four distinct regions of the sky, coinciding with well-known legacy fields.  Each region contains two adjacent shallow fields, and in two of the four regions there is also an adjacent deep field.  

The prefix for each DES field name is derived from the name of the legacy field in which it is located:  `X' for the fields lying in the XMM-LSS footprint, `C' for the fields clustered around the Chandra Deep Field - South (CDFS), `E' for the fields in and around Elais-S1, and `S' for fields located in SDSS - Stripe 82.  The centroids of each field are given in Table~\ref{tbl:fields}.  Shallow fields have a suffix of 1 or 2, with the more northerly field given the designation 1.  Deep fields have a suffix of 3.  All fields in the same region contain a small (order $1$\%) amount of overlap with one another.

\begin{deluxetable}{ l c  c  c }
\tablewidth{0pt}
\tabletypesize{\scriptsize}
\tablecaption{Field Locations\label{tbl:fields}}
\tablehead{
  \colhead{Legacy Field} &
  \colhead{DES Field} &  
  \colhead{RA (J2000)} & 
  \colhead{DEC} 
}
\startdata
\multirow{3}{*}{CDFS} & C1 & 03h 37m 05.83s  &  -27:06:41.8 \\
                      & C2 & 03h 37m 05.83s  &  -29:05:18.2 \\
                      & C3 & 03h 30m 35.62s  &  -28:06:00.0 \\
\hline
\multirow{2}{*}{Elais-S1} & E1 & 00h 31m 29.86s  &  -43:00:34.6 \\
                          & E2 & 00h 38m 00.00s  &  -43:59:52.8 \\
\hline
\multirow{2}{*}{SDSS Stripe 82} & S1 & 02h 51m 16.80s  &   00:00:00.0 \\
                                & S2 & 02h 44m 46.66s  &  -00:59:18.2 \\
\hline
\multirow{3}{*}{XMM-LSS} & X1 & 02h 17m 54.17s  &  -04:55:46.2 \\
                         & X2 & 02h 22m 39.48s  &  -06:24:43.6 \\
                         & X3 & 02h 25m 48.00s  &  -04:36:00.0
\enddata
\end{deluxetable}

\begin{figure}[!t]
	\centering
	\includegraphics[angle=0,width=3.5in]{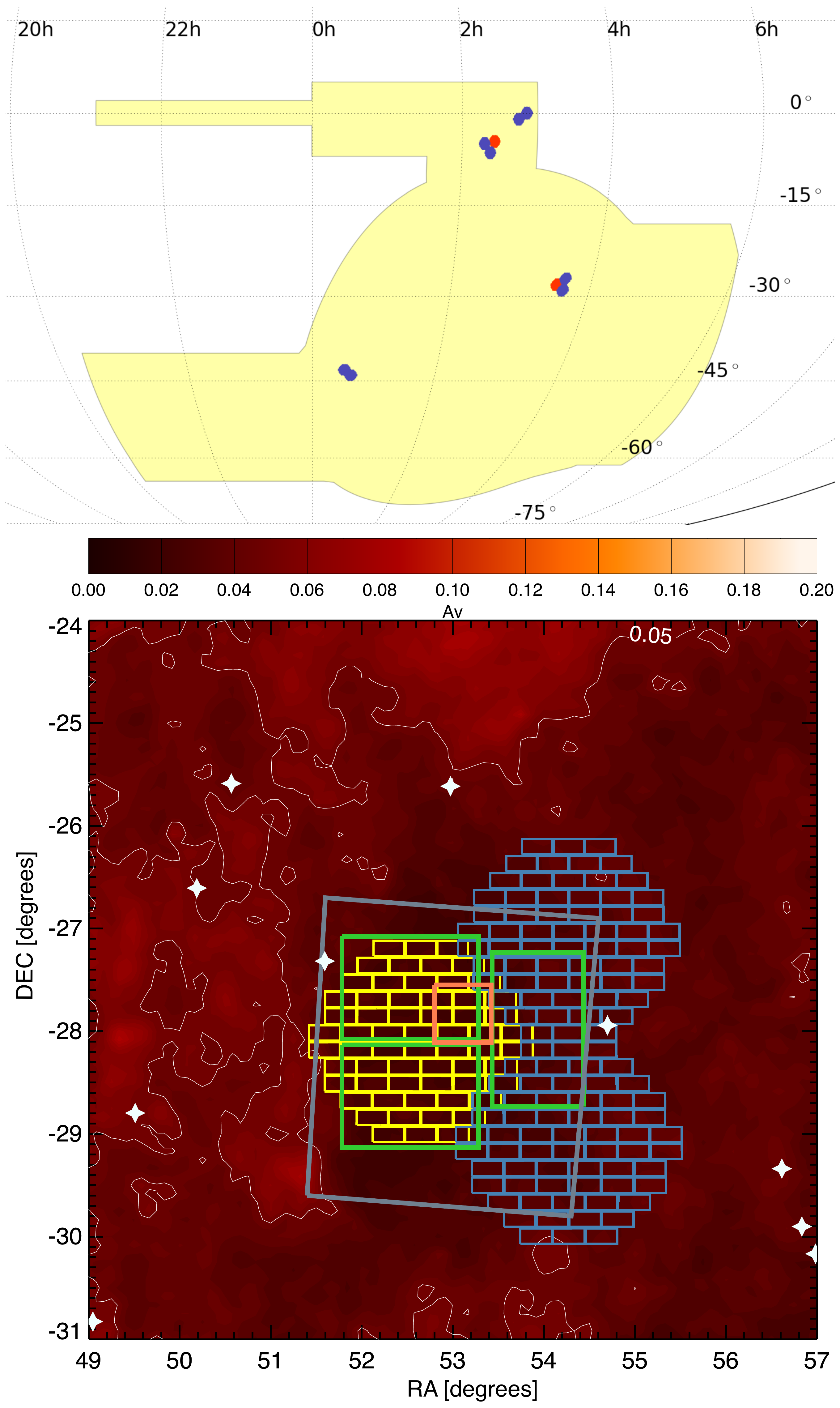}
	\caption{\textbf{Top:} The DES footprint (yellow), with the DES-SN shallow (blue) and deep (red) fields overplotted.  A Mollweide equal-area projection is used.  Positions of fields are listed in Table~\ref{tbl:fields}.  \textbf{Bottom:} Enlarged view of the DES-SN C fields.  The shallow fields (C1 and C2) are again in blue, and deep C3 is outlined in yellow.  These DECam footprints are plotted over a contour plot of MW extinction.  Bright stars ($M_V<8$ mag) are plotted in white. Overplotted are the boundaries of fields observed by SWIRE (grey), VIDEO (green), and CDFS (orange). 
	\label{fig:footprint}}    
\end{figure}

The DES-SN fields lie within the DES-wide footprint to benefit from a consistent photometric calibration (Figure~\ref{fig:footprint}).  This constraint forces all fields to be relatively close in RA, spanning only three hours.  The fields are broadly distributed in declination in order to allow for spectroscopic follow-up observations at low airmass from northern observatories for half of the fields (X and S), while the C and E fields are more southerly to allow for longer windows at low airmass and better avoidance of the Moon.  

In addition to considerations of calibration and spectroscopic follow-up, the DES-SN fields were located with Milky Way extinction and ancillary data in mind.  Low-extinction regions were preferentially selected ($E(B-V)<0.02$ mag in 8 of 10 fields), and overlap with other surveys was optimized.  Particular attention was paid to overlap with deep near-infrared \citep[VIDEO; ][]{Jarvis2013} and mid-infrared \citep[SWIRE; ][]{Lonsdale2003} survey regions.  Finally, field centroids were adjusted to minimize the area lost from masking of bright stars and their bleed trails.

\subsection{Observing Strategy}
\label{sec:obsstrat}
DES observes in 5-6 month seasons, starting in mid-late August and ending in early-mid February.  The season is constrained from being extended by the requirement of low-airmass observations in the compact DES-wide footprint.  The first and last month of each season are primarily scheduled as half-nights to ensure footprint visibility.  The length of the continuous observing season helps to minimize edge effects for DES-SN light curves, particularly for highly time-dilated SNe at $z\ge1$.  Most DES nights are dark, but CTIO schedules grey and bright time for DES as well to provide the community time with DECam at a range of RA and sky brightness conditions.  As such, the structure of DES-SN observations---survey duration, cadence, sky brightness---are the result of these competing interests.

DES uses an algorithmic scheduler \citep[{\tt ObsTac};][]{Neilsen2014} to determine the survey program (DES-SN or DES-wide), field, and filter to observe given the present observing conditions, the completeness of the DES-wide footprint, and the length of time $\Delta t_{\textrm{seq}}$ since the last accepted observation of each DES-SN sequence.  A `sequence' is defined as a series of exposures that are not interrupted once they begin, regardless of changing conditions.  The number of exposures per filter, per field is listed in Table~\ref{tbl:tobs}.  For each of the DES-SN shallow fields a sequence is all of the exposures in all of the filters (\emph{grizz}), while each filter-epoch is treated as a distinct sequence in each DES-SN deep field (\emph{ggg},\emph{rrr},\emph{iiiii},\emph{zzzzzzzzzzz}).  This uncoupling of the long deep-field filter-epochs introduces scheduling flexibility for {\tt ObsTac} to better optimize observations.

A DES-SN sequence is triggered if $\Delta t_{\textrm{seq}}\ge4$d and the seeing (measured at zenith in the $i$-band) is $\ge 1.1\arcsec$, or with no lower limit on the seeing if $\Delta t_{\textrm{seq}}\ge7$d.  Priority is always given to the sequence with the largest $\Delta t_{\textrm{seq}}$.  There are also upper limits on the projected seeing for the deep ($1.3\arcsec$) and shallow ($1.8\arcsec$) fields to minimize poor-quality data.  DES-SN observations require the predicted sky brightness for a filter-epoch (in mag/arcsec$^2$ above dark) to be less than 3/3/2/2 for $g/r/i/z$.  For the shallow fields this is loosened to 5/4/2/2 if $\Delta t_{\textrm{seq}}\ge7$d, as otherwise the filters being tied together in one observing sequence would result in long gaps \emph{in red filters} due to the brightness of the moon.  {\tt ObsTac} additionally requires a starting airmass $< 1.5$ per sequence, though this is loosened at the edges of each season.

The data quality (DQ) for each exposure is assessed based on an analysis of its output from the difference imaging pipeline \citep[{\tt DiffImg};][]{Kessler2015}.  There are three possibilities for the status of an exposure:  Pass, Fail, or Junk.  Pass means minimal acceptable DQ has been achieved, and Fail means it has not.  Specifically, an image fails DQ if the measured point-spread function (PSF), converted to $i$-band zenith, is $>2.0\arcsec$, or if the artificial sources of magnitude 20 we insert into our pipeline have a measured signal-to-noise ratio (SNR) of $<20$ ($<80$ for the deep fields).  Junk means that the pipeline was unable to process the image, either due to instrumental errors or exceedingly poor weather.  If an exposure is labeled as Fail or Junk, then it is not considered `accepted', and the clock for re-taking the sequence ($\Delta t_{\textrm{seq}}$) is not reset to zero.

\subsection{Survey Summary}
\label{sec:SurveySummary}

The DES-SN program took 6,877 exposures totaling 487.69 hours of on-sky time during the first three years of the survey.  Data quality was assessed as Pass for $87.7$ percent of the exposure time.  The mean number of total (Pass) epochs per field, per season was 29.3 (24.6) for each shallow field and 25.6 (22.7) for each deep field.  The number of good filter-epochs per season varied from 15-28, while the total number ranged from 21-32.

The mean duration of the observing season for each DES-SN field was 167 days, with only small variations across the seasons (163/168/170).  The observing season per field varied from 153 days (X3$r$ in Y3) to 182 days (C3$z$ in Y2).  Typically the southern fields have a longer continuous visibility than the more northernly fields (171 days for SN-C and SN-E; 160 and 164 for SN-S and SN-X, respectively.)  

The mean cadence for DES was 7.4d when considering only good-quality imaging; the cadence was 6.1d when including all imaging.  The shallow fields have a slightly better cadence than the deep fields (7.3d and 7.8d, respectively).  The shallow fields vary between a cadence of 7.1d to 7.5d, and in the deep fields -- where bands are observed independently -- there is no effective difference in cadence (range of 7.7d to 7.9d).  The median for all of the above quantities is 7d.  

\begin{figure*}[htp]
	\centering
	\includegraphics[width=\textwidth]{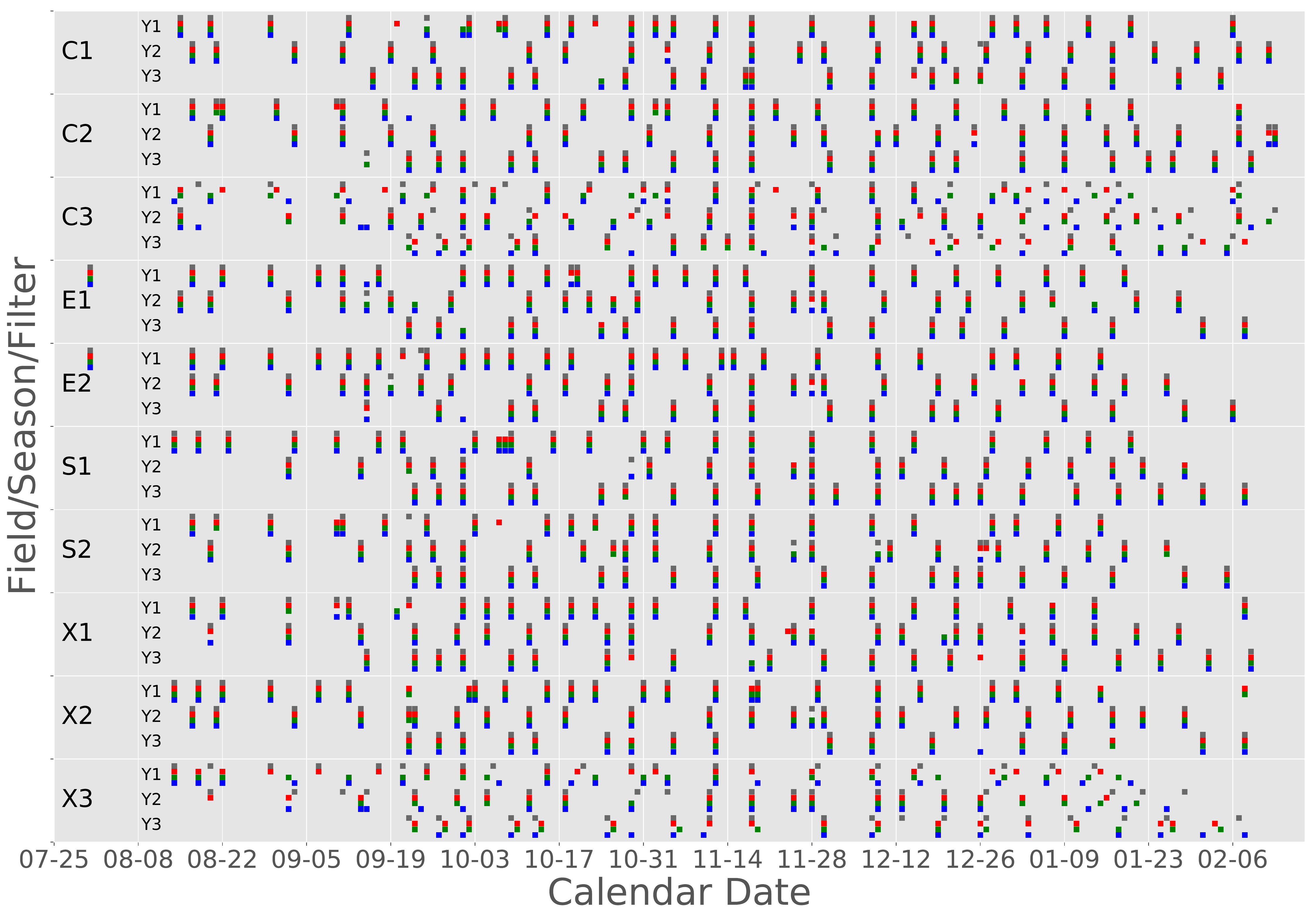}
	\caption{Every accepted observation taken over the first three seasons of DES-SN, with $g/r/i/z$ denoted as blue/green/red/grey.  Each of the 10 fields is grouped together and data are shown for all three seasons.  In the deep fields individual filters can be observed on different nights, whereas the shallow fields are grouped together as one observing block.  
    \label{fig:CadenceFull}}
\end{figure*}

\begin{figure}[htp]
	\centering
	\includegraphics[angle=0,width=3.5in]{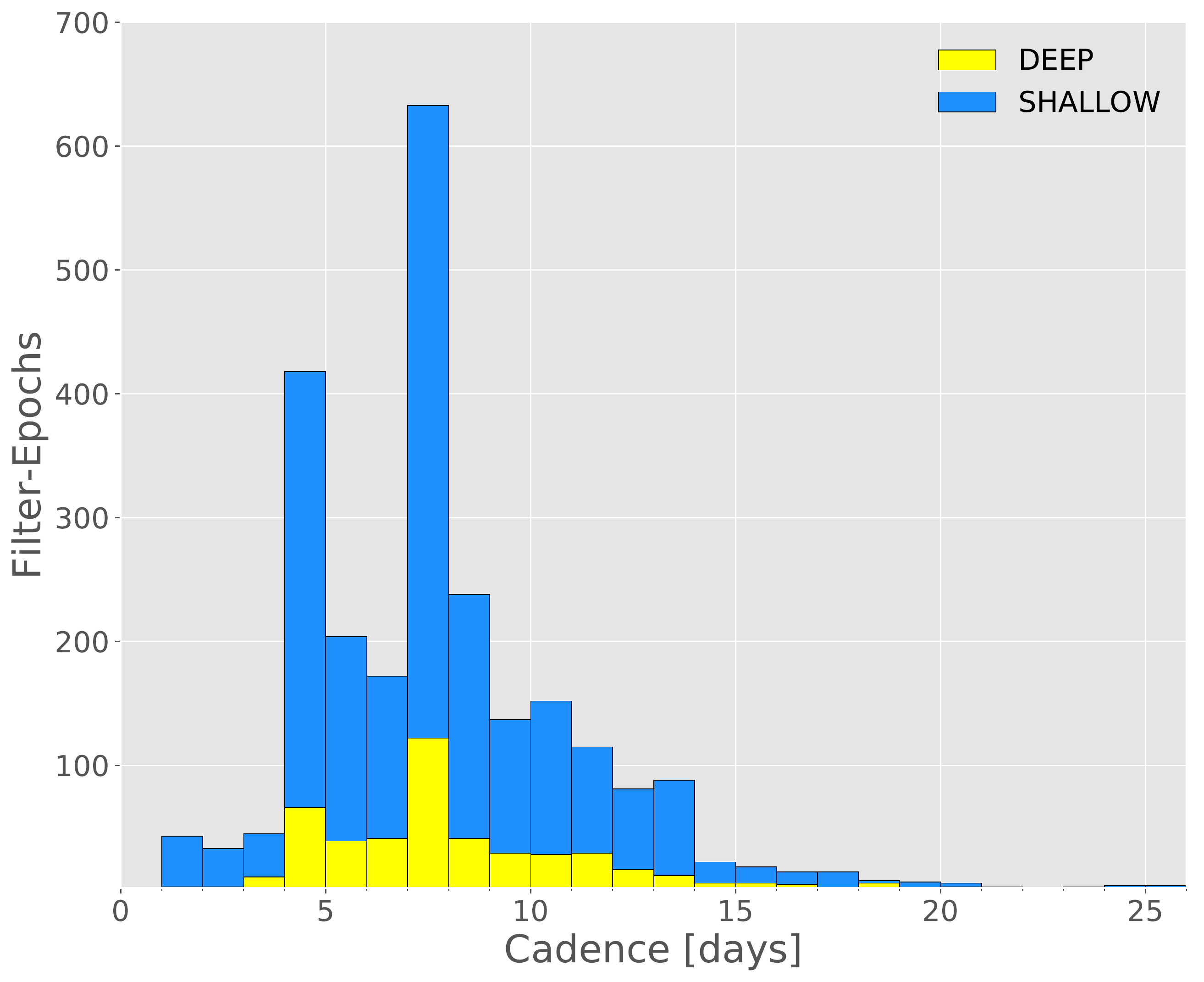}
	\caption{Stacked histogram of the number of days between good-quality data in a single filter-epoch over the first three years of the DES-SN program, split between deep (yellow) and shallow (blue) fields.  Note the peak at 4 days (when observations can be made owing to adequate seeing), and at 7 days (when observations begin to be forced regardless of the seeing).
    \label{fig:CadenceHist}}
\end{figure}

We show the full observing history of the first three years of DES-SN in Figure~\ref{fig:CadenceFull}, and condense this information into a histogram in Figure~\ref{fig:CadenceHist}.  68\% of all epochs were taken with a cadence of 4-8 days, and 23\% were taken with a cadence of 9-13 days.  Although observations are given top priority programatically at $\Delta t_{\textrm{seq}}=7$d, a number of factors cause a long tail to higher cadences.  In particular, there are poor weather nights, nights DES is off-sky for community time, nights when sky brightness is above the observing threshold, and nights when the time allotted for programmed DES-SN sequences exceeds the time of field visibility.

We note that there are cases where we have a cadence $<4$ days, seemingly in contradiction with {\tt ObsTac}.  There are two causes for this.  The majority of the short-cadence exposures (79\%) come from a shallow-field sequence where one filter fails DQ but the others pass; in these cases $\Delta t_{\textrm{seq}}$ is not reset, which can lead to short cadences for some filters.  The remaining short-cadence exposures are due to data-processing lags, where DQ was not accessed prior to the next night's observations.  

We have only 61 filter-epochs with a gap in the cadence $>15$ days, on average 0.5 per filter-epoch per season.  Fewer than half of these gaps occur between September 15 and January 31.  Therefore, most of the large light-curve gaps are at the very beginning (when weather is often poor), or at the very end of the season, when overriding of {\tt ObsTac} was permitted.  As the end of the DES-SN season is for completing already discovered SN light curves rather than searching for new SNe, assessment of overall DES collaboration needs permitted a reduced DES-SN cadence without sacrificing SNe that could potentially be part of a cosmological-analysis sample.

\begin{figure}[htp]
	\centering
	\includegraphics[angle=0,width=3.5in]{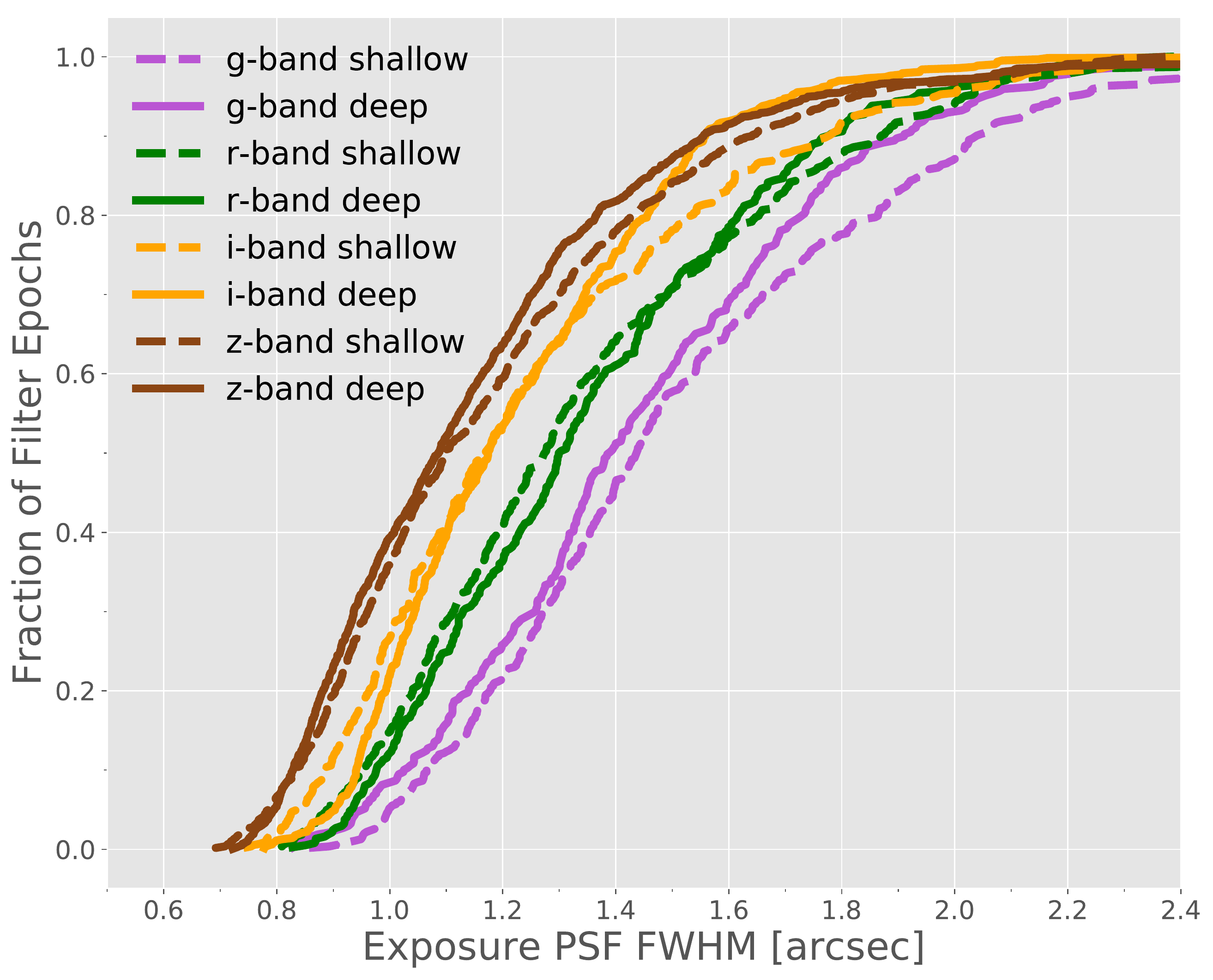}
	\caption{Distribution of observed PSF FWHM (in arcseconds) for each exposure in the first three years of the DES-SN survey.
    \label{fig:psf}}
\end{figure}

In Figure~\ref{fig:psf} we show the cumulative distribution function (CDF) for the measured full-width at half-maximum intensity (FWHM) of the PSF for each DES-SN exposure, split by band and depth.  The measured FWHM is worse for bluer bands, as the atmosphere produces a larger PSF for smaller wavelengths. The median observed FWHM in $griz$ (in arcseconds) is 1.41/1.29/1.17/1.09.  We note that {\tt ObsTac} does not use this measurement but rather the $i$-band zenith PSF to schedule observations, and as expected from our observing algorithm, the median of this statistic is consistent amongst all bands.  We also note that the distributions in Figure~\ref{fig:psf} are similar between deep and shallow fields, with the exception that the poorest $\sim10$\% of images in the deep field were taken in significantly better conditions than those in the shallow fields, which is also to be expected from our observing algorithm.

\begin{figure}[htp]
	\centering
	\includegraphics[angle=0,width=3.5in]{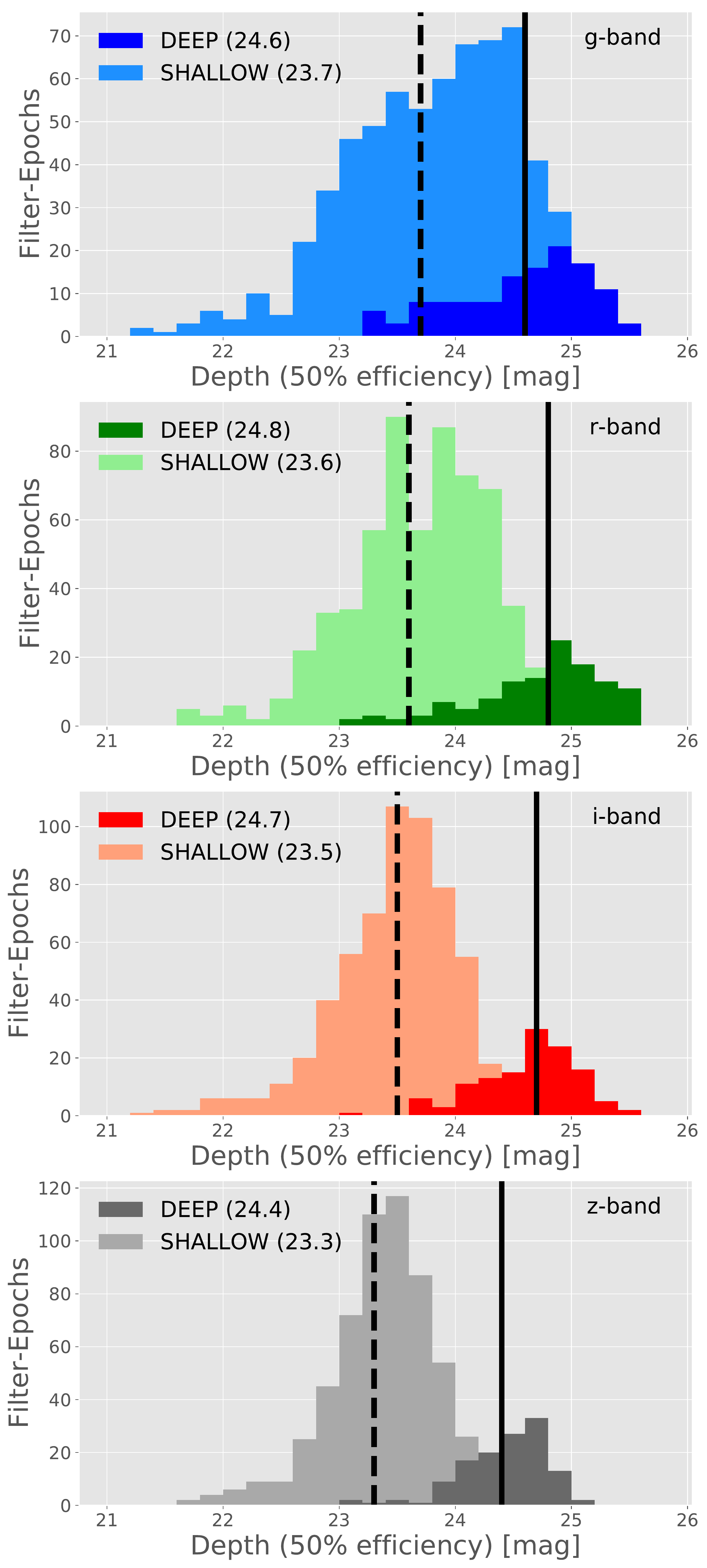}
	\caption{Depth of the DES filter-epochs in the first three seasons of the survey, split by deep and shallow fields.  Median depth in the deep and shallow fields are shown by solid and dashed lines, respectively.  
    \label{fig:depth}}
\end{figure}

In Figure~\ref{fig:depth} we plot the limiting magnitude of each filter-epoch, split by band.  The median depth across all bands is very similar, which was the intended outcome of our chosen exposure times.  We note that despite our usage of the term `shallow', these fields have per-exposure depths of $\sim 23.5$ mag, deeper than SDSS and equal to or slightly deeper than Pan-STARRS Medium-Deep Survey \citep{Sako2018,Rest2014}.  The larger variation in bluer bands is an effect of observing across a large variety of sky brightness conditions.

\section{Transient Identification}
\label{sec:PhotCands}

DES-SN data are processed nightly to find new transient candidates and to update the photometry for previously identified candidates.  All data processing for DES takes place at the National Center for Supercomputing Applications (NCSA\footnote{\url{http://www.ncsa.illinois.edu}}) in Urbana-Champaign, Illinois.  Here we briefly describe these steps and how they have evolved during the survey.  We end this section with statistics for detections over the first three seasons of DES-SN.  

\subsection{Image Processing \& Difference Imaging Pipeline}
\label{sec:DiffIm}

DES Observations are transferred from CTIO to NCSA within minutes of the end of each exposure, where they are preprocessed by the DES Data Management team (DESDM).  This step includes bias subtraction, flat fielding, bad-pixel masking, and corrections for crosstalk and nonlinear pixel response.  Next the images are run through the Supernova Single Epoch (SNSE) pipeline, where saturated stars and their bleed trails, cosmic rays, and satellites are masked out, and the sky is measured and subtracted from the image.  Full details of preprocessing and SNSE pipelines can be found in \citet{Morganson2018}.

Transients are identified from these images via the DES-SN Difference Imaging Pipeline {\Diff}, described in detail by \citet{Kessler2015}.  Much of {\Diff} makes use of publicly available codes or modifications thereof.  All exposures in a given filter-epoch are coadded to form a single `search image'.  Template images for a given season are created by coadding images taken in good conditions during previous DES observing seasons (during Y1, images from SV were used as templates).  Source extraction from both template and search images allows for a common astrometric solution to be determined.  The images are PSF-matched, the template is subtracted from the search image, and sources are extracted from the resulting `difference image'.  Every source detected by {\Diff} on every filter-epoch is classified as an `object', and is saved in a database.  These steps are run on each CCD individually.   

\subsection{autoScan}
\label{sec:ML}

We subsequently evaluate each object to determine whether it is a real astronomical point source or an artifact of the reduction pipeline (unmasked artifact, subtraction error, etc.).  For this we developed a supervised machine-learning algorithm \citep[{\tt autoScan};][]{Goldstein2015} that assigns a score (0-1) to each object, where a higher score indicates a higher likelihood of the object being a non-artifact.  The score is a function of many features computed directly from the images, most of which are ways to quantify the shape, size, and pixel-level flux distribution within the object.  \citet{Goldstein2015} determine that the {\tt autoScan} score should be $\ge0.5$ to optimize the detection efficiency with a minimal false detection rate.  An object with an {\tt autoScan} score $\ge0.5$ is labeled an `ml\_object' to specify that it has passed machine learning and is most likely a detection of a real astrophysical transient.

\subsection{Candidates and Transients}
\label{sec:Cands}

The presence of multiple spatially and temporally coincident `objects' triggers the creation of a `candidate'.  This is the most basic level in DES-SN of defining a likely astrophysical transient.  We define spatially coincident as objects detected in different images within $1\arcsec$ of the same position, both of which have an {\tt autoScan} score $\ge0.3$, in any combination of filters and detected less than 30 days apart.  We note the loose {\tt autoScan} threshold is intended to minimize the number of real transients that are discarded at this stage.  All candidates receive a unique SuperNova IDentification number, or SNID. PSF-fitted photometry (i.e., `forced photometry') is measured on all previous images at the candidate's position and updated with each subsequent observation \citep{Kessler2015}.

We define a `transient' (hereafter \emph{transient}) as a candidate consisting of at least two ml\_objects and detected on more than one night.  This higher threshold simultaneously removes slow-moving asteroids (which may be spatially coincident over one night) and lowers the contamination rate, a necessity for spectroscopic follow-up.   We run our host-matching algorithm (Section~\ref{sec:Hosts}) and do real-time light-curve fitting (Section~\ref{sec:psnid}) on all \emph{transients}.

All \emph{transients} receive unique names of the format `DESXXYYzzzz' based on their location and the time of their discovery.  For a \emph{transient} discovered in a given observing season, `XX' represents the last two digits of the year in which that observing season began: 13, 14, and 15 for DES Y1, Y2, and Y3, respectively.  The DES-SN field the \emph{transient} was first discovered in is given by `YY', and `zzzz' is a unique alphabetical identifier within the season that ascends as the season progresses (a, ..., z, aa, ..., zz, etc.).  Owing to multiple reprocessings of data as incremental improvements were made to {\Diff}, the naming system is no longer strictly sequential, and some irregularities -- such as names no longer associated with a \emph{transient} -- occur.  However, in no circumstance was a name ever used for two different candidates.

Finally, we developed a `transient\_status' flag to further improve the efficiency of our spectroscopic follow-up.  This empirically derived flag removes the most commonly found cases of artifacts that pass our simple \emph{transient} criteria.  There are four indicators of an artifact we evaluate for: (i) pixel correlation (repeated detections on a single pixel, column, or row); (ii) band multiplicity (non-physical fraction of all detections occurring in single band); (iii) large temporal separation between detections ($\Delta$t between ml\_objects); and (iv) large quantity of poor subtractions (indicating multiple good ml\_objects by chance).  These criteria are evaluated for each \emph{transient} separately over each DES season; a \emph{transient} passing all criteria is given a positive flag.  If a transient has a positive flag for multiple seasons, the flags are added together, and these multi-season transients (MSTs) are removed from the list of potential spectroscopic follow-up targets.  Objects with a positive transient\_status flag in one observing season are labeled as single-season transients (SSTs), and form the set of candidates of likely SNe that become possible targets for spectroscopic follow-up programs.

\subsection{Host Galaxies}
\label{sec:Hosts}

Host galaxies are assigned to candidates via the Directional Light Radius method (DLR; \citealp{Gupta2016}).  The DLR method uses the {\tt SExtractor} model-independent shape parameters $A$, $B$, and $THETA$ derived from the second moments of the observed galaxy light distribution.  Using these {\tt SExtractor} parameters, we determine the distance between the candidate and the galaxy in normalized units of the light profile projected in the direction of the transient ($d_{\textrm{DLR}}$).  This quantity is computed for all galaxies within a $15\arcsec$ radius.  The closest galaxy in this dimensionless measure is assigned as the likely host, provided $d_{\textrm{DLR}}\le4$.  If no galaxy satisfies this criterion, the candidate is considered to be hostless.  We note that for all DES candidates with an identified host galaxy via the DLR method, the same galaxy would have been selected as the likely host in 98.8\% of cases if we simply chose the galaxy with smallest angular separation from the candidate.

We do host-galaxy matching using the DES SVA1-GOLD galaxy catalog, created from DES Science Verification data, for DES-SN host galaxies.  Although deeper catalogs can be created from subsequent data, SVA1-GOLD has the advantage of using the same catalog for all DES-SN seasons and being free of contaminating SN light.  Furthermore, these catalogs are complete to $r\approx24.4$ mag ($r\approx25.5$ mag in the deep fields) which, as will be described in Section~\ref{sec:Spec}, means they are more than adequate for the purposes of our live-SN and host-galaxy spectroscopic programs.

\subsection{Photometric Classification}
\label{sec:psnid}

During the DES observing seasons we run the Photometric Supernova IDentification software \citep[\PSNID;][]{Sako2011} for every active candidate.  \PSNID\ compares the light curve for each candidate to a grid of templates of the most common SN subtypes (\SNIa, type II SNe, and type Ib/c SNe), measuring the best-fit parameters for each of the models to the data.  It then computes the probability that the model describes the data (`FITPROB'), and uses this to determine the Bayesian probability of the candidate being a particular subtype (`PBAYES').  This information is updated with each new epoch of photometry until the end of the observing season in which the candidate was discovered.  The results of \PSNID\ help shape our spectroscopic SN-follow-up program (Section~\ref{sec:Spec}) up to and just past peak brightness for the candidate; afterward, these preliminary typings contribute to target-selection for the host-galaxy spectroscopic follow-up.  We run \PSNID\ in two modes:  both without any priors and with a photo-$z$ prior from the host galaxy where one has been identified.  We used the DESDM neural network photo-$z$ catalog for these galaxies, which is described by \citet{Sanchez2014}.

\subsection{Statistics Summary}
\label{sec:ProcStats}

In Table~\ref{tbl:stats} we present statistics describing the quantities of objects, candidates, and \emph{transients} found in each of the first three seasons of DES.  We discovered 12,015 single-season \emph{transients} over the first three seasons of DES-SN, an average of 24 per night (see mean observing season duration in Section~\ref{sec:SurveySummary}).

The differences in statistics between seasons are quite small, with ${\cal O}(10\%)$ variation among seasons.  We note that candidates observed across multiple seasons---e.g., active galactic nuclei (AGN)---are attributed to the first season they appear in.  As such, Y1 has a much higher number of candidates than the other seasons, but a consistent number of SSTs.
 
Our requirement of multi-epoch detection for all spectroscopic targeting, when coupled with the DES-SN cadence of $>6$ days, limits our ability to obtain early-time follow-up spectra of transients.  However, the value of the multi-epoch requirement for cleanly removing asteroids from our \SNIa\ follow-up sample can be clearly demonstrated.  The DES-S and DES-X fields have the greatest proximity to the ecliptic, where asteroids are most likely to be detected.  In S1+S2 the number of candidates per field was 250\% that of the further-removed shallow fields (C1, C2, E1,and E2); for X1+X2 the rate was 192\%.  However, all shallow field pairs have the same number of \emph{transients} to within 5\%, as expected once asteroids are removed.   

\begin{deluxetable*}{ll | lll | ll}[!htbp]
\tablewidth{0pt}
\tabletypesize{\scriptsize}
\tablecaption{DES-SN Detection Statistics\label{tbl:stats} }
\tablehead{
  \colhead{Type\tablenotemark{1}}  &
  \colhead{Total} & 
  \colhead{Y1}    &
  \colhead{Y2}    &
  \colhead{Y3}    &   
  \colhead{Deep\tablenotemark{2}} &
  \colhead{Shallow\tablenotemark{2}}
}
\startdata
Objects\tablenotemark{a} & 4.88M  &  1.63M  &  1.66M   &  1.59M  & 265K & 137K \\
ML Objects\tablenotemark{b} & 1.21M  &  421K   &  389K    &  397K   & 52,623 & 37,139 \\
Candidates\tablenotemark{c} & 45.9K  &  18,489  &  13,586   &  13,836  & 1100 & 1626 \\
Transients\tablenotemark{d} & 17,215 &  6404   &  5521    &  5290   & 731 & 527 \\
SSTs\tablenotemark{e} & 12,015 &  4059   &  4326    &  3630   & 593 & 347 \\
\enddata
\tablenotetext{1}{All types are defined in Section~\ref{sec:Cands}.}
\tablenotetext{2}{Mean quantity per field type, per season.}
\tablenotetext{a}{Detection in a single filter-epoch by {\tt DiffImg}.}
\tablenotetext{b}{Objects with ({\tt autoScan} score $\ge0.5$).}
\tablenotetext{c}{Spatially coincident detections.} 
\tablenotetext{d}{Candidates consisting of ML Objects on multiple epochs.} 
\tablenotetext{e}{Transients with `transient\_status' $>0$ in one season only.}
\end{deluxetable*}

\section{Spectroscopy}
\label{sec:Spec}

In this section we give a full overview of the supernova spectroscopy program for the first three seasons\footnote{We do not discuss in this paper our spectroscopic SN follow-up program during the SV season.  Follow-up resources were limited, and the DES observing season was abbreviated with low cadence and data-quality.  None of these data are included in the DES-SN3YR analysis.} of DES-SN. First we describe in Section~\ref{sec:SpecStrategy} the multi-pronged strategy for the live SN follow-up program and the importance of each individual component.  We detail all spectroscopic observations taken of DES SN candidates, sorted by observatory and including a description of the selection criteria used, in Section~\ref{sec:SpecData}.  We describe our spectroscopic reduction and classification methods in Section~\ref{sec:SpecAnalysis}, and present our final sample of classified DES SNe in Section~\ref{sec:SpecSummary}.  In Section~\ref{sec:ssf} we will take the observations described in this Section and turn to the question of deriving the effective spectroscopic selection function of the survey.

\subsection{Supernova Follow-up Strategy \& Target Selection}
\label{sec:SpecStrategy}

The differentiating characteristic of DES-SN when compared to all previous \SNIa\ programs is that the primary cosmological analysis is designed for a sample of photometrically classified \SNeIa, allowing for a more efficient usage of the survey data than relying solely on spectroscopic confirmations.  However, spectroscopy is still vital for DES-SN, but with a different set of priorities compared to previous surveys.  Rather than our final sample consisting solely of transients classified as \SNeIa\ via our spectroscopic program, the photometrically classified sample will be shaped by the redshifts and classifications obtained via spectroscopy. 

The spectra collected as part of our \SNIa\ follow-up program serve several purposes.  They constitute a truth sample for training photometric classification in DES; they are used for further analysis of detailed properties, such as the correlation between intrinsic color and velocity \citep{Foley2011a,Foley2011b}; and they provide a sample of \SNeIa\ in low-luminosity galaxies that would otherwise be missed by our spectroscopic follow-up program of SN host galaxies, allowing a study of systematics due to correlations between \SNeIa\ properties and the mass of their host galaxies \citep{Sullivan2010,Kelly2010,Lampeitl2010}.  Most crucially, though, they allow for a cosmological analysis in the traditional mold of a spectroscopically confirmed \SNIa\ sample, testing the quality of our data and our analysis techniques.  Here we describe the different modes of our spectroscopy program, and how we implemented them in order to achieve these goals. 

\subsubsection{Host-Galaxy Spectroscopy}
\label{sec:SpecHost}

The focus of this paper is the \emph{live} spectroscopy of SNe in DES, but due to its outsize importance we briefly describe here the DES-SN host-galaxy spectroscopic follow-up program.  

Live follow-up of SNe not only provides a spectroscopic classification of the transient, but the redshift of the object being characterized as well.  For precision measurements of cosmological parameters with photometrically classified \SNeIa, DES-SN requires spectroscopic redshifts for these transients.  Obtaining spectroscopic redshifts from host galaxies as opposed to the SN spectra themselves is much more efficient: the observations are not time-critical, the source density increases with time, and the targets can be repeatedly observed to obtain higher SNR and depth.  Thus, the primary source of redshifts for DES-SN transients is from their host galaxies.

Most of these redshifts are obtained via the `OzDES' survey (PI C. Lidman).  OzDES began in 2013 with the primary goal of obtaining the host-galaxy redshifts necessary for DES \SNeIa\ cosmology.  This is a 100-night program, spread over 6 years, using the 3.9m Anglo-Australian Telescope (AAT).  There are 392 fibers on the AAOmega/2dF multi-fiber spectrograph covering a field of view that is nearly identical to the footprint of DECam.  This allows us to obtain spectra of tens of thousands of targets across the DES-SN fields, repeatedly observing galaxies over multiple observing runs until sufficient SNR is built up to obtain a secure redshift.   

Minimal cuts are placed on selecting candidates for host-galaxy follow-up: it must have an observed peak during a DES observing season, peak SNR$>5$ in at least two different filters, and be a single-season transient (as defined in Section~\ref{sec:Cands}).  The likely host galaxies for these transients are identified, and those with a magnitude $r<24$ (measured within the $2\arcsec$ fiber diameter of 2dF) are selected.  A full description of the OzDES observing strategy, including target classes, a list of observations, and the public spectroscopic catalog can be found in \citet{Yuan2015} and \citet{Childress2017}.

We note that not all host-galaxy redshifts were obtained by the OzDES survey.  OzDES assembled a custom redshift catalog in the DES-SN fields from an exhaustive literature search, which enabled the survey to more efficiently allocate its fibers by avoiding galaxies with already known spectroscopic redshifts.  Additionally, live SN follow-up is often able to obtain a spectroscopic redshift from the host galaxy.  However, in this latter case we still observe the host galaxy with AAT once the SN has faded to obtain a galaxy spectrum free of contaminating SN light.

\subsubsection{Magnitude-Limited Sample}
\label{sec:SpecMagLim}

We now describe the first of our three live follow-up programs of \SNeIa.  A magnitude-limited sample is a useful component for a spectroscopic supernova survey in that it creates an easily quantifiable selection function; if the resources allowed, we would carry out magnitude-limited follow-up to the full depth of our survey.  We thus created a Magnitude-Limited program to characterize the brightest SN candidates in DES in an otherwise unbiased manner.  No \PSNID\ probabilities are used in this selection; all SSTs detected by DES are eligible to be targeted with this program.  This is to later allow for testing and validation of photometric-classification routines.

The primary source of observations for this program was also the OzDES program, as it is efficient for targeting objects with a low source density (Section~\ref{sec:SpecHost}) and is rapidly configurable, allowing the target list to be updated in real time.  On any given OzDES observing night, all active SSTs with $r<22.7$ or $i<22.7$ mag in the field being observed have a fiber placed on them.  These observations were supplemented by follow-up with other observatories---primarily the Southern African Large Telescope (SALT) and MMT---to obtain classifications of SSTs not obtained at the AAT owing to weather, observing cadence, or other classification inefficiencies.  The goal for completeness in this campaign was all SSTs brighter than $r = 22$ mag, though as we will demonstrate later this goal was not achieved.

\subsubsection{Faint Hosts}
\label{sec:SpecFaintHost}

Since the inclusion of a SN in the photometric cosmology analysis requires a spectroscopic redshift, all SNe occurring in galaxies too faint for our OzDES host-galaxy spectroscopy campaign ($r>24$ mag) would be excluded from our cosmology sample.  This creates a selection bias against \SNeIa\ as a function of both decreasing host-galaxy mass as well as increasing redshift.  To characterize this bias we carried out a follow-up program for spectroscopically classifying \SNeIa\ in faint host galaxies, ensuring that these \SNeIa\ can be retained in our final analysis.  The redshift information obtained from the spectrum can be used alone or in conjunction with the spectroscopic classification.  

\SNIa\ candidates for this program are selected based on their early-time light curve with \PSNID, and prioritized based on the apparent faintness of their host-galaxy.  There is overlap between this program and the Magnitude-Limited one, as a bright candidate can also be hostless.  In these cases resources from either campaign can be used to secure a spectroscopic type.  The Faint Hosts program (which targets SNe in faint galaxies, not the galaxies themselves) had dedicated observing time at the Very Large Telescope (VLT) and the Gran Telescopio Canarias (GTC), and additional data were taken with Keck and Magellan. 

\subsubsection{Representative Sample}
\label{sec:SpecRep}

The last of the three live \SNIa\ follow-up programs in DES-SN is designed to obtain a representative sample: a spectroscopically confirmed sample of \SNeIa\ that evenly samples the redshift distribution of SNe in the final photometrically classified analysis.  This sample has many important uses, such as allowing us to test for environmental dependence and color evolution of spectroscopic properties with redshift.  Quantifying the effect of evolution in spectroscopic properties on a representative sample is likely to be helpful in our future DES cosmology analysis of a photometrically classified \SNIa\ sample, which will lack this spectroscopic information.  Having a representative sample of SNe is also important as a training set for future methods of photometric classification that rely on machine learning.

For this follow-up program, we first determine likely \SNIa\ candidates using \PSNID\ fits.  The other follow-up programs independently fill portions of this parameter space; e.g., low-redshift SNe for the Magnitude-Limited sample and (preferentially) high-redshift SNe for the Faint Host sample.  As a result, the Representative Program primarily observes \SNeIa\ at redshifts $0.3 < z < 0.7$, and is biased toward higher-mass hosts at high redshift, areas that are missed by our other follow-up programs.  Dedicated observing programs at Gemini and Magellan comprise the majority of this sample, though data for this was collected at Keck and MMT as well.  The final sample for this program should be thought of as a subset of data from all live SN programs.

\subsubsection{Non-Ia Supernovae}
\label{sec:SpecNonIa}
DES is a cosmology survey, and thus DES-SN has been designed to discover, measure, and confirm \SNeIa.  But as a deep and wide transient survey, there are many other interesting types of transients that can be found in the data.  We briefly note here three additional classes of transients for which we have made a concerted effort to obtain follow-up spectroscopy: superluminous supernovae (SLSNe), tidal disruption events (TDEs), and type II supernovae (SNe\,II).

These programs are much smaller than our \SNIa\ program, with the only follow-up time specifically allocated for such observations coming from the VLT (SLSNe), Magellan (SN\,II), and Gemini (SN\,II).  However, since SLSNe occur preferentially in very low-mass galaxies, we have also obtained spectral confirmations of these objects from our faint-host program.  We have also unintentionally classified SLSNe with other \SNIa\ dedicated programs \citep[DES15E2mlf;][]{Pan2017}, but no \SNeIa\ were classified from observations of potential SLSN targets.  Although non-Ia SNe were observed as part of the DES-SN follow-up program, further discussion and publication of these spectra are reserved for papers analyzing these data \citep{Papadopoulos2015,Smith2016,Pan2017,Smith2018}.

\subsection{ATC}
\label{sec:ATC}

To successfully carry out our multiple spectroscopic observing programs across a globally distributed collection of telescopes requires real-time coordination, long-term transient monitoring, and centralized data storage.  For this purpose we developed a tracking database and web application called ATC, hosted at the National Energy Research Scientific Computing Center (NERSC).

Information about each DES transient, including coordinates, photometry, discovery date, and host-galaxy association, are used to seed an initial portfolio.  This portfolio develops over time as additional photometry is acquired, with light-curve fits and SN subtype probabilities continuously updated using \PSNID.  Spectroscopic follow-up is coordinated and scheduled through ATC, as tags are applied to portfolios to indicate which transients should be observed, when, and from where.  Finder charts from DECam imaging are generated on demand and made available to observers via the web.  After follow-up, tags are updated and observing reports attached to the ATC.  Reduced spectra are uploaded for inspection and science use by the collaboration.  Redshifts and classifications derived from the spectra are recorded in ATC for each transient.  All of the DES-SN spectroscopic follow-up, which we detail in the following section, was dependent on the ATC.

\subsection{Data}
\label{sec:SpecData}

\begin{deluxetable*}{ l l l l l l l l l l }
\tablewidth{0pt}
\tabletypesize{\scriptsize}
\tablecaption{DES-SN Spectroscopy Program\label{tbl:specPrograms}}
\tablehead{
  \colhead{Observatory} &
  \colhead{PI\tablenotemark{a}} & 
  \colhead{Instrument} &  
  \colhead{Wavelength [nm]\tablenotemark{b}} & 
  \colhead{Allocation\tablenotemark{c}} & 
  \colhead{Season} & 
  \colhead{Spectra\tablenotemark{d}} & 
  \colhead{Spec Ia\tablenotemark{e}} & 
}
\startdata
AAT\tablenotemark{f,g}    & C. Lidman     & 2dF/AAOmega    & 380-880  & 48n    & Y1,Y2,Y3 & 1002 & 77 \\
AAT\tablenotemark{g}      & C. Smith      & 2dF/AAOmega    & 380-880  & 2n     & Y1       & 7    & 1  \\
Gemini\tablenotemark{g,i} & R. Foley      & GMOS           & 520-990  & 18h   & Y1       & 2    & 2  \\
Gemini\tablenotemark{i,j} & R. Foley      & GMOS           & 520-990  & 39.6h  & Y3       & 25   & 18 \\
Gemini\tablenotemark{i}   & L. Galbany    & GMOS           & 520-990  & 10h    & Y3       & 5    & 1  \\
GTC                       & F. Castander  & OSIRIS         & 480-920  & 54.6h  & Y1,Y2,Y3 & 19   & 10 \\
Keck\tablenotemark{g}     & A. Filippenko & DEIMOS         & 455-960  & 4n     & Y1,Y2,Y3 & 7    & 2  \\
                          &               & LRIS        & 340-1025 & 11.5n + 26h & Y1,Y2,Y3 & 18 & 9 \\
Magellan                  & R. Kirshner   & LDSS3          & 425-1000 & 2n     & Y2,Y3    & 16   & 5  \\
                          &               & IMACS          & 390-1000 & 8n     & Y2,Y3    & 41   & 28 \\
Magellan                  & S. Gonz\'alez-Gait\'an & LDSS3 & 425-1000 & 4n     & Y3       & 23   & 9  \\
Magellan                  & D. Scolnic    & LDSS3          & 425-1000 & 1n     & Y3       & 8    & 7  \\
MMT                       & R. Kirshner   & BCS            & 330-850  & 7n     & Y2,Y3    & 31   & 12 \\
SALT                      & M. Smith      & RSS            & 385-820  & 41.59h & Y1,Y2    & 21   & 6  \\
SALT                      & E. Kasai      & RSS            & 385-820  & 37.5h  & Y3       & 31   & 12 \\
VLT\tablenotemark{f,h}    & M. Sullivan   & X-Shooter      & 300-2480 & 14.1n  & Y2,Y3    & 89   & 47 \\
VLT\tablenotemark{h}    & M. Sullivan   & X-Shooter      & 300-2480 & 12h    & Y3       & 7    & 2 
\enddata
\tablenotetext{a}{Program IDs for the spectroscopic campaigns listed here can be found in the acknowledgements.}
\tablenotetext{b}{Wavelength range given is representative of the typical instrumental setup and may vary in a given program.}
\tablenotetext{c}{Actual on-sky time for each program is less than the allocation, depending on weather conditions; the priority level assigned to ToO programs; and the availability of suitable targets.  Hours are used for queue-scheduled time, nights for classical time.}
\tablenotetext{d}{Number of spectra obtained for DES transients by the program.  Does not refer solely to \SNIa\ candidates.}
\tablenotetext{e}{Discovery spectra only.  Some programs obtained classifiable \SNIa\ spectra for previously classified objects; we do not double-count those here.}
\tablenotetext{f}{Allocation listed is part of a long-term program that continues into subsequent seasons.}
\tablenotetext{g}{Program includes DES \SNIa\ targets, but is not the primary purpose of the program.}
\tablenotetext{h}{Data from the NIR arm not used in most observations.}
\tablenotetext{i}{Observations made with Gemini-South}
\tablenotetext{j}{Observations made with Gemini-North.}
\end{deluxetable*}

Here we describe the observing campaigns undertaken by DES as a function of observatory, including the number of candidates observed, mode of observations, and resulting classifications from each telescope.  In Table~\ref{tbl:specPrograms} we list the main details of the spectroscopic programs for which DES \SNIa\ targets were observed.  We note that total allocated time, not on-sky time, is listed here.  The fraction of time used on DES-SN targets varies depending on weather, observatory down time, and other programs sharing the same time allocation.  Additionally, the number of spectra is the total number of observations, including repeated observations of the same target and non-Ia SNe.  Thus, a comparatively low number of \SNIa\ classifications for a given program does not equate to a low classification efficiency.  A description of the classification procedures will be given in Section~\ref{sec:SpecAnalysis}.

\begin{deluxetable*} {  l cc ll c lll lr }
\tablecaption{Spectroscopic Observing Log:  DESY1-Y3\label{tbl:specAll}}
\tablewidth{0pt}
\tabletypesize{\small}
\tablehead{
\colhead{Transient} & \colhead{Telescope}  & \colhead{Instrument} &
\colhead{Date} & \colhead{Date} & \colhead{Exposure} &  
\colhead{Seeing}    & \colhead{Airmass}    & \colhead{Slit\tablenotemark{a}} &
\colhead{Observed\tablenotemark{b}}   & \colhead{\% Flux\tablenotemark{c}} \\ 
\colhead{Name}          & \colhead{}           & \colhead{} &
\colhead{[UT]}      & \colhead{[MJD]}    & \colhead{Time [s]}   &  
\colhead{[arcsec]}  & \colhead{}  & \colhead{[arcsec]} &
\colhead{Mag$_i$}          & \colhead{Increase}
}
\startdata
DES13C1c        &  AAT      &   AAOmega/2dF  &  2013-10-01    &  56566.66     &  3x2400,1x1800          &  2.5-5   &  ....   &  2.0     & 22.3 [0.6] &  116.0 \\ 
DES13C1d        &  AAT      &   AAOmega/2dF  &  2013-10-01    &  56566.66     &  3x2400,1x1800          &  2.5-5   &  ....   &  2.0     & 23.7 [0.6] &  9.79 \\ 
DES13C1e        &  AAT      &   AAOmega/2dF  &  2013-10-01    &  56566.66     &  3x2400,1x1800          &  2.5-5   &  ....   &  2.0     & 21.9 [-7.5] &  50.0 \\ 
DES13C1eie      &  AAT      &   AAOmega/2dF  &  2013-11-30    &  56626.71     &  1x2400                 &  NA      &  ....   &  2.0     & 24.1 [1.4] &  7.43 \\ 
DES13C1eie      &  AAT      &   AAOmega/2dF  &  2013-12-01    &  56627.65     &  2x2400,1x1596          &  1.6     &  ....   &  2.0     & 24.1 [0.4] &  7.43 \\ 
DES13C1feu      &  SALT             &      RSS       &  2013-10-08    &  56573.91     &  1x2400            &  1.2     &  1.18    &  1.5     & 19.9 [1.3] &  140.0 \\ 
DES13C1feu      &  AAT      &   AAOmega/2dF  &  2013-10-30    &  56595.68     &  2x2400                 &  1.3     &  ....   &  2.0     & 20.9 [-5.3] &  31.8 \\ 
DES13C1feu      &  AAT      &   AAOmega/2dF  &  2013-11-02    &  56598.69     &  2x2400                 &  1.6     &  ....   &  2.0     & 21.2 [3.4] &  24.1 \\ 
DES13C1feu      &  AAT      &   AAOmega/2dF  &  2013-11-30    &  56626.71     &  1x2400                 &  NA      &  ....   &  2.0     & 22.0 [1.4] &  11.6 \\ 
DES13C1feu      &  AAT      &   AAOmega/2dF  &  2013-12-01    &  56627.65     &  2x2400,1x1596          &  1.6     &  ....   &  2.0     & 22.0 [0.4] &  11.6 \\ 
DES13C1feu      &  AAT      &   AAOmega/2dF  &  2013-12-26    &  56652.63     &  1x2400                 &  1.3     &  ....   &  2.0     & 22.7 [0.5] &  6.07 \\ 
DES13C1fpp      &  AAT      &   AAOmega/2dF  &  2013-10-30    &  56595.68     &  2x2400                 &  1.3     &  ....   &  2.0     & $>$21.9 [-5.3] &  $<$19.4 \\ 
DES13C1gki      &  AAT      &   AAOmega/2dF  &  2013-10-30    &  56595.68     &  2x2400                 &  1.3     &  ....   &  2.0     & 22.9 [-1.5] &  260.0 \\ 
DES13C1gol      &  Keck             &      DEIMOS    &  2013-10-06    &  56571.53     &  1x1200            &  0.6       &  ...     &  ...     & 22.5 [3.7] &  117.0 \\ 
DES13C1hwx      &  AAT      &   AAOmega/2dF  &  2013-10-30    &  56595.68     &  2x2400                 &  1.3     &  ....   &  2.0     & 22.6 [-1.5] &  885.0
\enddata
\tablenotetext{a}{For AAOmega/2dF, fiber diameter is given in place of slit width.}
\tablenotetext{b}{Apparent magnitude from DES-SN observation on the epoch $t_{\textrm{phot}}$ closest to the time of spectroscopic followup $t_{\textrm{spec}}$.  In brackets we give the value of $t_{\textrm{phot}}-t_{\textrm{spec}}$.}
\tablenotetext{c}{Brightness of target at time of observation relative to the surface brightness of the background.}
\tablecomments{This table is available in full online as part of the DES-SN3YR data release: \urlDR}
\end{deluxetable*}

In Table~\ref{tbl:specAll} we present our observation log of spectra taken of DES transients (full table is available online).  For each observation we list the DES transient name; the telescope, instrument, exposure time, and setup; the MJD and UT date of the observation; the seeing and airmass; the magnitude of the transient at the time of observation; and the \% increase of the flux from the transient over the background flux.  We note the majority of the table consists of OzDES follow-up; there are over 1000 spectra of transients taken as part of OzDES, and 343 spectra from all other telescopes in our program combined. 

Except where otherwise noted, all spectra described herein have been reduced using standard routines with IRAF (Image Reduction and Analysis Facility\footnote{IRAF is distributed by the National Optical Astronomy Observatory, which is operated by the Association of Universities for Research in Astronomy (AURA) under a cooperative agreement with the National Science Foundation.}).  Basic data calibration (bias/overscan subtraction, flux calibration, and wavelength calibration) was performed by the individual observing teams. Supernova spectra were extracted from the geometrically corrected two-dimensional (2D) spectra, often with significant amounts of host-galaxy background; for higher-redshift targets embedded in their hosts, modest spatial-width apertures were used in extracting the supernova spectra to minimize host contamination.

\subsubsection{Anglo-Australian Telescope (AAT)}
\label{sec:telAAT}

The majority of spectroscopic observations of DES transients have been performed as part of the OzDES Survey \citep[Section~\ref{sec:SpecHost};][]{Yuan2015,Childress2017}.  OzDES uses the 2dF fiber positioner on the 3.9\,m AAT telescope at Siding Spring Observatory in Australia.  The rapidity of the AAT fiber configuration software allows for same-day updates to the target list.  We placed active transients at highest priority (i.e., over-riding all other classes of target allocations) into each observation.  In Y1 and Y2 all active transients with $r < 22.5$ mag were added to the OzDES observing queue.  In Y3 this was extended to objects with $i<22.5$ mag (for any $r$), as well as rising transients (brightened by $>0.3$ mag between detections) that were brighter than $23.0$ mag in either $r$ or $i$.  In practice, a complete sample of bright transients was not obtained, as the time between observing runs (typically $>1$ month) and weather prevented all transients from being observed with any magnitude cutoff.  However the biases are still minimal, as the follow-up campaign was not influenced by any inferred supernova properties (e.g., color, light-curve shape, host properties).  

AAT observations form the majority of our Magnitude-Limited sample described in Section~\ref{sec:SpecMagLim}, as well as the Host-Galaxy program detailed in Section~\ref{sec:SpecHost}.  Observations were made using the AAOmega dual bench spectrograph with the 2dF fiber positioners \citep{Saunders2004,Smith2004}.  The AAOmega setup consisted of the x5700 dichroic with 580V (blue) and R385 (red) gratings, with a resulting continuous wavelength coverage of $3800-8800$\,\AA.  Both CCDs were upgraded between Y1 and Y2, resulting in better quantum efficiency in both arms, as well as fewer cosmetic defects in the blue and less fringing in the red.  The 2dF instrument places 392 science fibers, each with a diameter of 2$\arcsec$, over a 2-degree diameter field.  The spectra were reduced with a modified copy of v6.2 of the 2dfdr pipeline \citep{Croom2004}, which is further described by \citet{Childress2017}.

Over the course of the first three DES observing seasons (Y1/Y2/Y3) OzDES was allocated 48 nights (12/16/20); a further two nights allocated in Y1 to another DES AAT program (PI C. Smith) were integrated into OzDES, both operationally and in terms of data reduction.  We obtained 1009 spectra (235/307/467) of 533 (127/180/226) distinct transients\footnote{In early versions of the {\Diff} pipeline artifacts were not efficiently vetoed.  This led to candidates targeted by OzDES in Y1 which were later identified as artifacts.  For clarity we include here numbers referring only to real transients.}.  For over half of the transients (274) we obtained a single spectrum, while for approximately half of the remainder (128) we obtained multiple spectra in one observing run.  Thus, while most transients (402) were observed in only one run, 99 were observed in exactly two runs, and 32 were observed in three or more runs.  The typical gap between observing runs was approximately one month.

The total numbers of supernovae spectroscopically confirmed with OzDES are 78 \SNeIa\ (12/26/40), 14 SNe\,II (3/3/8), and 1 SN\,Ic (1/0/0).  The median redshift of \SNeIa\ classified by OzDES is $z=0.279$.  Spectroscopic classification efficiency with this program is very low owing to the observational setup: 2$\arcsec$ diameter fibers mean more host-galaxy light than SN light typically enters the fiber, and our follow-up strategy does not take into account the increase over background surface brightness.  Unlike our other spectroscopic follow-up programs, transients were not removed from the OzDES queue once they possessed a spectroscopic classification.  Therefore, we have classification-quality spectra from OzDES for an additional 36 transients which we do not consider as `classified' by OzDES. 

\subsubsection{Gemini}
\label{sec:telGEM}

We obtained 32 spectra of 29 DES transients using the Gemini Observatory; 18 spectra taken at the 8.1\,m Gemini North telescope on Maunakea, Hawaii, and 14 spectra taken at the 8.1m Gemini South telescope on Cerro Pachon, Chile.  Observations were carried out in long-slit mode with the Gemini Multi-Object Spectrograph \citep[GMOS;][]{Hook2004}, one of which is present on each telescope.  With the exception of two spectra in Y1, all of our observations presented here were made in Y3, after the upgrade to Hamamatsu CCDs on GMOS-S (2014A), but prior to the upgrade on GMOS-N (2017A).  All but one target were observed with the R400 grating, which yields $\lambda/\Delta\lambda\ = R\approx1900$.  Order-blocking filter OG515 was used in most observations, yielding a useful wavelength range of $5200-9900$\,\AA, and the standard slitwidth was 1.0\arcsec.  Full details for each spectrum can be found in Table~\ref{tbl:specAll} online.  We note that all data with Gemini were taken in queue mode.

Gemini was the primary resource used for our Representative sample, though one of our programs was also part of the Non-Ia program.  We classified 21 of 29 objects targeted as \SNeIa\ (median $z=0.565$), 2 as SNe\,Ibc, and one as a SLSN.  All transients were initially targeted as likely \SNeIa\, including the high-$z$ SLSN DES15E2mlf \citep{Pan2017}, which was also the only transient observed multiple times with Gemini.

\subsubsection{GTC}
\label{sec:telGTC}

We obtained 19 spectra of 18 DES transients using the Optical System for Imaging and low-Intermediate-Resolution Integrated Spectroscopy \citep[OSIRIS;][]{Cepa2003} on the 10.4\,m GTC located at the Observatorio del Roque de Los Muchachos in La Palma.  All observations were made in long-slit mode using the R500R grism, yielding $R \approx 600$ with a useful data range of $4800-9200$\,\AA.  All but one observation used a 0.8\arcsec\ slit.  Observations took place over all three seasons, with 5/9/5 targets in DES Y1/Y2/Y3.  All observations were carried out in queue mode. 

We primarily used GTC for our Non-Ia and Faint Host programs.  We classified 10 \SNeIa\ (median $z=0.398$) and 5 SLSNe with GTC.    The remaining observations were re-observations of previously classified targets (one \SNIa, one SLSN, and one TDE).  Only one target was not classified.  

\subsubsection{Keck}
\label{sec:telKEC}

We obtained 25 spectra of 23 DES transients using instruments at the 10\,m W.~M.~Keck Observatory on Maunakea, Hawaii.  Seven spectra of distinct targets were obtained with the DEep Imaging Multi-Object Spectrograph \citep[DEIMOS;][]{Faber2003} on Keck II, and 18 spectra of 16 transients were followed up with the Low Resolution Imaging Spectrograph \citep[LRIS;][]{Oke1995} on Keck I.  DEIMOS observations utilized the LVMslitC slitmask, typically with a 1.0\arcsec\ slit, and a combination of the low-resolution 600ZD grating ($R\approx123$) with the GG455 long-pass order-blocking filter.  This provided a useful wavelength coverage of $4550-9600$\,\AA.  On LRIS the 600/4000 grism was used on the blue side and the 400/8500 grating was used on the red side, providing a wavelength coverage of $3400-10250$\,\AA.  Most observations used the 1.0\arcsec slit rotated to the parallactic angle to minimize the effects of atmospheric dispersion (\citealt{Filippenko1982}; in addition, LRIS has an atmospheric dispersion corrector).  Observations took place over all three seasons, with 4/12/9 targets in DES Y1/Y2/Y3.  All observations with Keck were taken as part of classically scheduled time.

Observations with Keck primarily were part of the Representative (17 targets) and Non-Ia (6) programs.  We classified 11 \SNeIa\ (median $z=0.443$), 1 SLSN, 1 SN\,II, and 1 SN\,IIn with Keck.  We took multiple spectra of a SLSN and a TDE.  Non-classifications were mainly due to poor observing conditions.  The data were reduced using standard techniques with routines written specifically for Keck+LRIS in the Carnegie {\tt python} ({\tt carpy}) package \citep{Kelson2000,Kelson2003}."

\subsubsection{Magellan}
\label{sec:telMAG}

We obtained 88 spectra of 86 transients in DES using the twin 6.5\,m Magellan telescopes at Las Campanas Observatory, Chile.  47 spectra were obtained on the Low Dispersion Survey Spectrograph (LDSS-3) on the Clay Telescope, while 41 were obtained with the Inamori Magellan Areal Camera and Spectrograph \citep[IMACS;][]{Bigelow1998} on the Baade Telescope.  IMACS observations were made in Short Camera Mode with the Gri-300-17.5 grism, providing wavelength coverage over $4250-9500$\,\AA\ with resolution $R\approx1100$.  Various slit widths as well as slit masks were used, though no observations were made in MOS mode.  LDSS-3 data were taken with three different observing programs, but all observations were made with the VPH-ALL grism, covering the wavelength range $4250-10000$\,\AA\ with $R=860$.  No observations with Magellan occurred in Y1, increasing to 31 and 57 in Y2 and Y3, respectively.  All observing time used was classically scheduled.  

Observations with Magellan were split amongst nearly all programs, with targets falling in the Representative, Magnitude-Limited, and Non-Ia samples.  We classified 49 \SNeIa\ (median $z=0.348$), 6 SN\,II, 1 SN\,Ibc, and 1 TDE with Magellan.  The only targets with multiple spectra were a previously classified SLSN, a TDE, and a \SNIa\ candidate previously observed in poor conditions.

\subsubsection{MMT}
\label{sec:telMMT}

We obtained 31 spectra of 28 DES transients using the Blue Channel Spectrograph \citep[BCS;][]{Angel1979} on the 6.5\,m MMT at the Fred Lawrence Whipple Observatory in Mount Hopkins, Arizona.  All observations are single-slit and utilized the 300GPM grating and a clear filter, with a resolution $R=740$ covering a wavelength range of $3300-8500$\,\AA.  Most observations used the 1.0\arcsec\ slit.  A majority of the spectra (22) were taken in Y2, with the remainder coming in Y3.  All observing time was classically scheduled.  

Observations with MMT were primarily part of the Magnitude-Limited sample, with a few targets in both the Representative and Non-Ia samples.  We classified 12 \SNeIa\ (median $z=0.302$), 1 SN\,II, and 1 SLSN-I with MMT.  Non-classifications were primarily due to poor weather and targets in the Magnitude-Limited sample having significant host-galaxy contamination.  The three targets observed twice were \SNIa\ candidates for which initial spectra were of low SNR.  

\subsubsection{SALT}
\label{sec:telSAL}

We obtained 52 spectra of 44 distinct DES transients using the Robert Stobie Spectrograph \citep[RSS;][]{Smith2006} on the 11\,m SALT at the South African Astronomical Observatory (SAAO).  All observations were made in long-slit mode using the pg0300 low-resolution grating ($R\approx350$) and the pc03850 UV order-blocking filter.  The extracted spectra cover a wavelength range of $3850-8200$\,\AA.  Most observations were carried out using a 1.5\arcsec\ slit, which is well matched for the typical seeing at the site.  We have observed DES transients with SALT in every DES season, with 10/6/28 targets in Y1/Y2/Y3, respectively.  All Observations were carried out in queue mode.  

Observations with SALT mainly belong to the Magnitude-Limited sample.  We classified 18 \SNeIa\ (median $z=0.175$), 3 SNe\,II, 2 SNe\,Ic, and 2 SNe\,Ibc with SALT.  This program had an additional goal of obtaining \SNIa\ spectra near maximum light for a detailed analysis project; this led to us obtaining spectra of \SNeIa\ that already had classifications.  Most repeat observations of a transient were due to the fixed zenith angle of the telescope, which results in a limited visibility window and led to us obtaining multiple observations to increase the effective SNR.  The only exception was DES15S2nr, a SLSN that was observed several times with SALT.  Data were reduced using PySALT\footnote{http://pysalt.salt.ac.za}, the SALT science pipeline \citep{Crawford2010}.

\subsubsection{VLT}
\label{sec:telVLT}

We obtained 96 spectra of 91 DES transients using the X-Shooter echelle spectrograph \citep{Vernet2011} on the 8.2\,m VLT at the European Southern Observatory (ESO) on Cerro Paranal, Chile.  X-Shooter has three arms (UVB, VIS, NIR), which combined provide continuous wavelength coverage over $3000-24800$\,\AA.  The resolution depends on the arm and slit-width used (typically $0.7-1.0$\arcsec), varying from $R\approx4000-10000$.  Most of our observations were in `stare' mode, as the magnitude range of our targets means only the UVB and VIS arms provide useful data.  However, for brighter targets we observed by nodding along the slit, which facilitates sky subtraction in the NIR arm.  Spectra were taken in Y2 (38) and Y3 (58), spanning ESO observing periods P93-P96.  We note that some of our observations took place before X-Shooter was moved from UT3 to UT2, after P93. Our larger VLT program (spanning 4 ESO semesters) was classically scheduled, while the smaller one was queue observing.

VLT was the primary source for our Faint Hosts program, though our queue program was targeted at SLSNe.  We classified 49 \SNeIa\ (median $z=0.541$), 10 SNe\,II, 2 SLSNe, and 1 SN\,Ic with VLT.  Multiple spectra were obtained for two candidates with initial low SNR observations, while time-series data were obtained for two non-Ia targets.  The data were reduced via a modified version of the {\tt EsoReflex} pipeline \citep{Freudling2013}, where the rebinning procedure on the highly dispersed echelle spectrum has been improved to obtain proper statistics for very low SNR data.  

\subsubsection{Additional Spectra}
\label{sec:telOther}

The programs above are supplemented by a small quantity of assorted additional spectra.  We obtained data with the Shane 3\,m telescope at Lick Observatory confirming DES14X2fna as a SN\,II, which later transitioned to a SN\,IIb.  We obtained spectra of DES14C3rap and DES14C1kia with the 4.1\,m Southern Astrophysics Research Telescope (SOAR) on Cerro Pachon, Chile; the former classified as a \SNIa, while the latter had been previously classified as a TDE.  We obtained via a DDT proposal a spectrum of DES13S2cmm using FORS2 at VLT, which became the first classification of a SLSN in DES \citep{Papadopoulos2015}.

Two DES transients were bright enough that classifications were made prior by other groups before we had the opportunity to obtain spectra of the candidates.  DES13C3avpi was classified as a \SNIa\ by the PESSTO collaboration using the 3.58\,m NTT \citep[as PS1-13eao;][]{Bersier2013}, and DES14S1rwf was also classified as a \SNIa, by the KISS collaboration using the 1.88\,m Okayama telescope \citep[as SN2014dy;][]{Matsumoto2015}. 

\begin{deluxetable*} {llllllr}
\tablecaption{Spectroscopically-Confirmed Supernovae in the DES-SN Three-Year Sample\label{tbl:specClass}}
\tabletypesize{\scriptsize}
\tablehead{
\colhead{Name} & 
\colhead{SNID} & 
\colhead{RA (J2000)} &
\colhead{DEC} & 
\colhead{$z$} & 
\colhead{$z_{\textrm{err}}$} &  
\colhead{Classification}
}
\startdata
DES13C1feu      &    1251839 &    53.266731 &   -26.964838 &  0.05982 &   0.0005 & SNIc \\ 
DES13C1hwx      &    1253039 &    54.418591 &   -27.527397 &    0.454 &    0.006 & SNIa \\ 
DES13C1juw      &    1253920 &    54.629311 &   -27.042770 &    0.196 &   0.0005 & SNIa \\ 
DES13C1ryv      &    1257366 &    53.899559 &   -26.969273 &   0.2118 &   0.0005 & SNIa \\ 
DES13C2acmj     &    1262128 &    53.267674 &   -29.469212 &   0.1137 &   0.0005 & SNIa \\ 
DES13C2bxd      &    1247673 &    54.510422 &   -29.157028 &  0.04042 &   0.0005 & SNIc \\ 
DES13C2dyc      &    1249851 &    55.218029 &   -29.399984 &   0.2159 &   0.0005 & SNIa \\ 
DES13C2jtx      &    1252955 &    54.722439 &   -28.774586 &    0.223 &    0.005 & SNII \\ 
DES13C3abhe     &    1262715 &    53.369957 &   -28.442709 &     0.69 &     0.01 & SNIa \\ 
DES13C3abht     &    1262214 &    53.502693 &   -28.660202 &     0.69 &     0.01 & SNIa 
\enddata
\tablecomments{This table is available in full online as part of the DES-SN3YR data release: \urlDR}
\end{deluxetable*}

\subsection{Transient Spectroscopic Classification}
\label{sec:SpecAnalysis}

We classify all reduced spectra using both {\tt SNID} \citep[SuperNova IDentification; v5.0;][]{Blondin2007} and {\tt Superfit} \citep[v3.5;][]{Howell2005} software.  These approaches use cross-correlation techniques ({\tt SNID}) and chi-squared minimization ({\tt Superfit}) to produce, for a given spectrum, a rank-ordered list of matches from a spectral library of supernovae, galaxies, or any other variable objects.  These codes also allow for external information, such as redshift and phase, to be included in the fit.  All fits were performed using the spectroscopic redshift obtained either from galaxy emission lines in the spectrum itself or from a custom redshift catalog created from an exhaustive literature search by the OzDES team where available.  The {\tt Superfit} classification results are given priority owing to its inclusion in the fitting process of galaxy contamination in the observed spectrum. 

Classifications were determined via visual inspection of the resulting rank-ordered fits by a subset of co-authors, and were based on consensus of the best-fitting templates. Spectra fall into one of the following broad categories: (i) a transient of some known type (see below), (ii) uncertain (spectrum contains some signal from a transient, but classification is unclear), or (iii) a galaxy spectrum with no obvious transient light. For the transient classification, we characterize the spectra as ``SNIa", ``SNIbc", ``SNII", ``SLSN-I", ``TDE", ``AGN", and ``M-star".  The TDE is an outlier, as this was classified by visual comparison with known TDE spectra without the use of {\tt SNID} or {\tt Superfit} results.  Further detailed subclassifications are not attempted in the analysis for this paper, as the typical DES spectrum is only intended to have a sufficient SNR to make broad classifications required for distinguishing \SNeIa\ from other types.  Peculiar subtypes of \SNeIa\ are mostly rejected in the light-curve fitting process, which will by necessity be the only method of rejecting these transients from the photometric-cosmology analysis as well.  Detailed subsets of SNe spectra with higher SNR will be analyzed in future DES papers.

Both the spectroscopic redshift and the phase of the SN, as determined from the light-curve evolution, must match the entire list of best-fitting templates for a definitive classification to be claimed.  If no spectroscopic redshift from the host galaxy is available, then the SN redshift can be fit as well, and a definitive classification can be claimed if the phase, type, and redshift are robustly determined by the fitting software.  

Where a classification is highly probable but with some (well-defined) uncertainty, we use the following classifications:  "SNIa?", "SNIbc?", "SNII?", and ``SLSN-I?".  These classifications are used in two cases.  The first case is when there is no independent host-galaxy prior on the redshift and a small fraction of viable fits exists with a redshift and/or type that differs from the primary classification.  The second case is where a spectrum is a good fit to templates over only one half of the spectrum for plausible reasons: a low-SNR spectrum, poor sky subtraction, or host-galaxy contamination (primarily in the red portion of the spectrum, as the SN is typically brightest in the blue).  The phase is still required to match the light curve in all cases.  We note that in Section~\ref{sec:SpecData}, for simplicity, we combined likely and definitive classifications.  We describe these in more detail in Section~\ref{sec:SpecSummary}.

\subsection{Spectroscopy Summary}
\label{sec:SpecSummary}

\begin{figure*}[htp]
  \centering	
  \includegraphics[width=\textwidth]{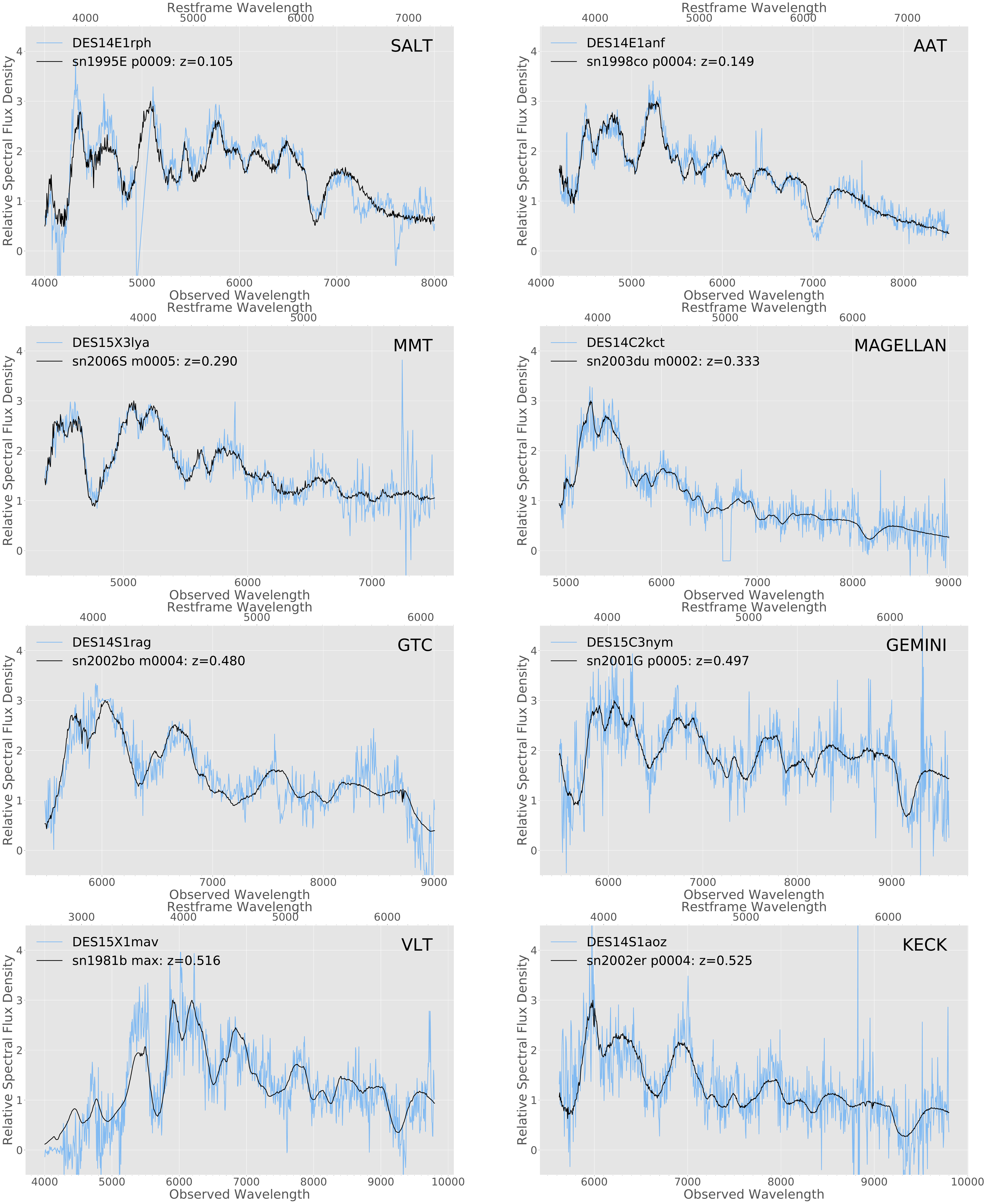}
  \caption{Observations from each of the observatories used for the DES-SN follow-up program.  Spectra are plotted in blue, with the best-fit supernova template overplotted in black.  Fits are derived using {\tt Superfit}, and host-galaxy contamination has been subtracted from the data.  All spectra shown here are classified as `SNIa'. \label{fig:Spectral_fits}}
\end{figure*}

In Table~\ref{tbl:specClass} we present our spectroscopically classified SNe from the first three seasons of the DES-SN program.  We have classified 307 SNe over a redshift range $0.017<z<1.86$ and a dynamic range of $>1000$ in peak observed flux.  We list totals for all certain and likely classifications in Table~\ref{tbl:classSum}.  Combining these, we have identified 251 spectroscopic \SNeIa, 34 SNe\,II (including IIn), 12 SLSNe, and 10 SNe\,Ib/Ic/IIb.  In Figure~\ref{fig:Spectral_fits} we present several of our spectroscopically classified \SNeIa\ with their best-fit template models overplotted.

\begin{figure}[htp]
	\centering
	\includegraphics[angle=0,width=3.5in]{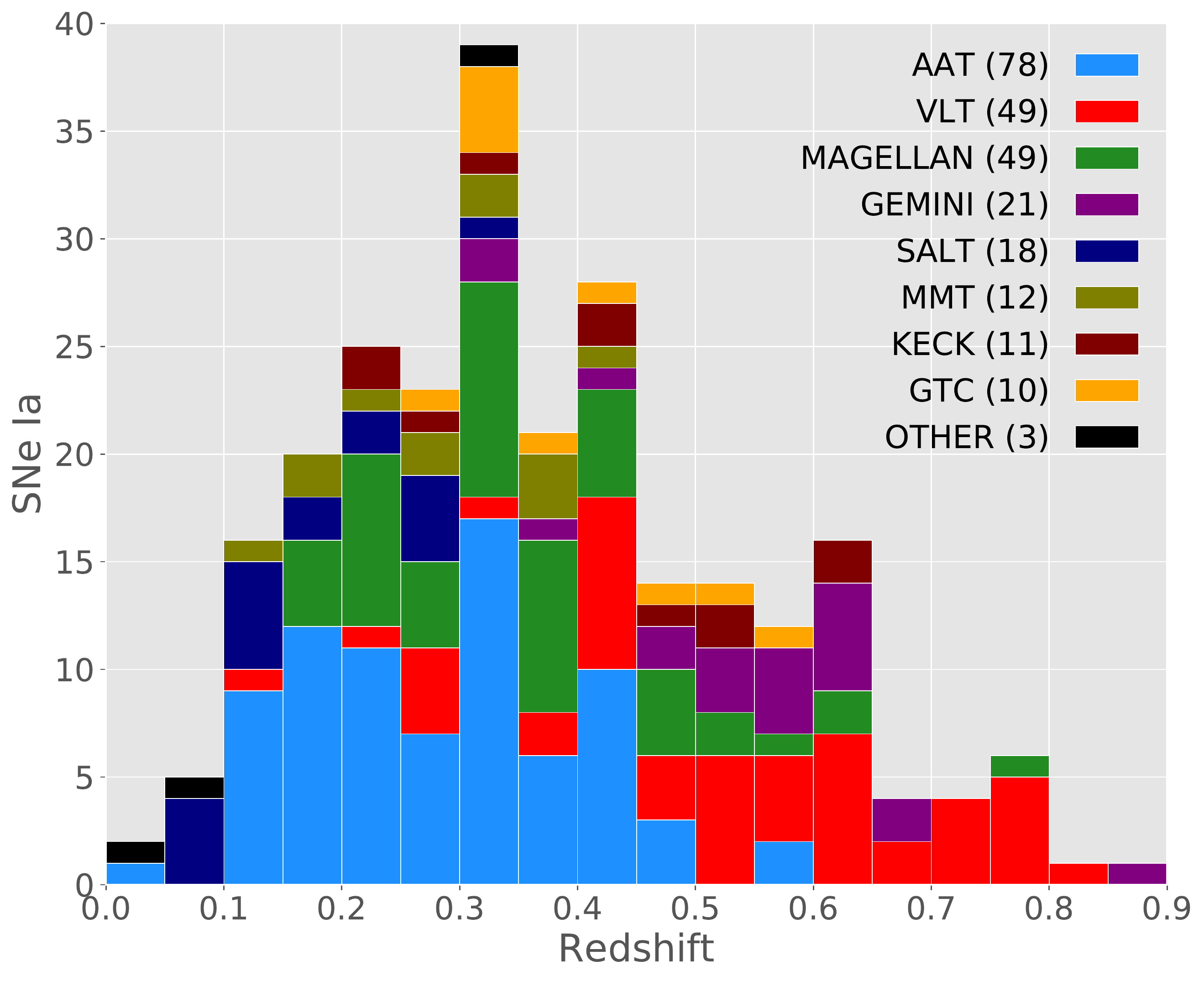}
	\caption{Redshift histogram of all 251 spectroscopically confirmed \SNeIa\ in the first three seasons of DES ($\Delta z=0.05$).  The histogram is split by discovery observatory, with the total quantity of \SNeIa\ classified by each telescope indicated in the legend.
	\label{fig:zHist}}
\end{figure}

In Figure~\ref{fig:zHist} we show the redshift distribution of the \SNeIa\ classified by our program, color-coded by the telescope which provided the classification.  The observing program that drives the follow-up for each telescope can be seen in the redshift range of classified \SNeIa\ in the figure.  VLT (Faint Hosts) dominates at high-$z$, while the AAT and SALT (Magnitude-Limited) fill out medium and low redshifts.  Magellan and Gemini extend from mid to high redshift owing to a mixture of Magnitude-Limited and Representative programs.  

\begin{figure*}[htp]
	\centering
	\includegraphics[width=\textwidth]{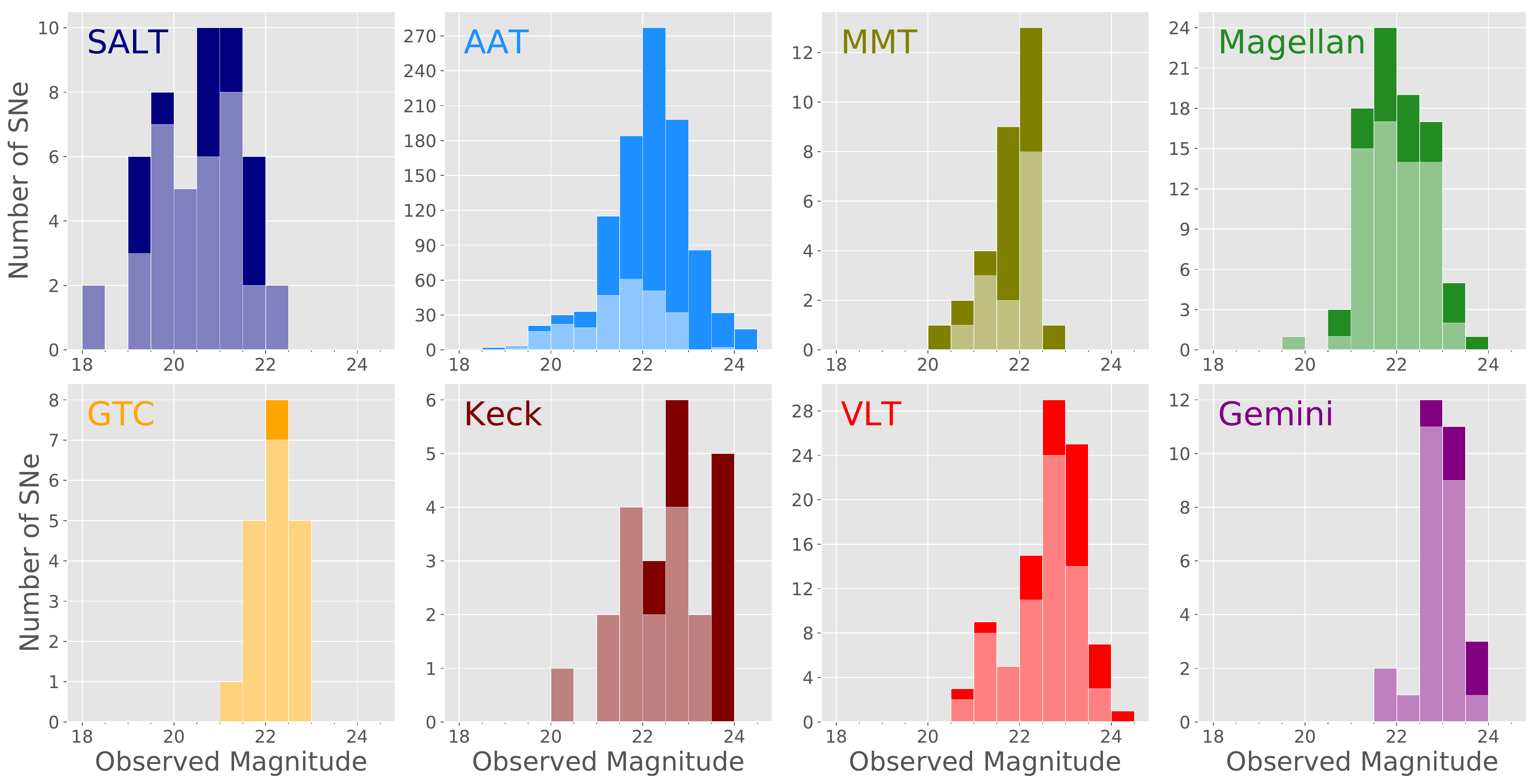}
	\caption{Observed apparent magnitude distributions for transients spectroscopically observed in the first three seasons of DES.  Magnitudes are $i$-band for all observatories other than AAT and SALT, which were $r$-band selected follow-up programs.  Distributions are shown separately for each telescope, in order of increasing median apparent magnitude: from $r=20.8$ for SALT to $i=23.0$ for Gemini. The lighter shaded histogram in each plot represents the subset of observations that resulted in a successful classification.  
	\label{fig:magHist}}
\end{figure*}

In Figure~\ref{fig:magHist} we show the apparent magnitude distribution of transients targeted, and classified, as a function of telescope, where the magnitude is taken from the DES epoch immediately preceding the spectroscopic observation.  Here we include \emph{all} classifiable spectra as successes, including where the transient targeted has previously been spectroscopically confirmed.  Note that while OzDES obtained nearly three times as many spectra as all the single-slit programs combined, the classification efficiency for AAT ($\sim25$\%) was far below that of the single-slit follow-up ($>70$\%).

\begin{figure}[htp]
	\centering
	\includegraphics[angle=0,width=3.5in]{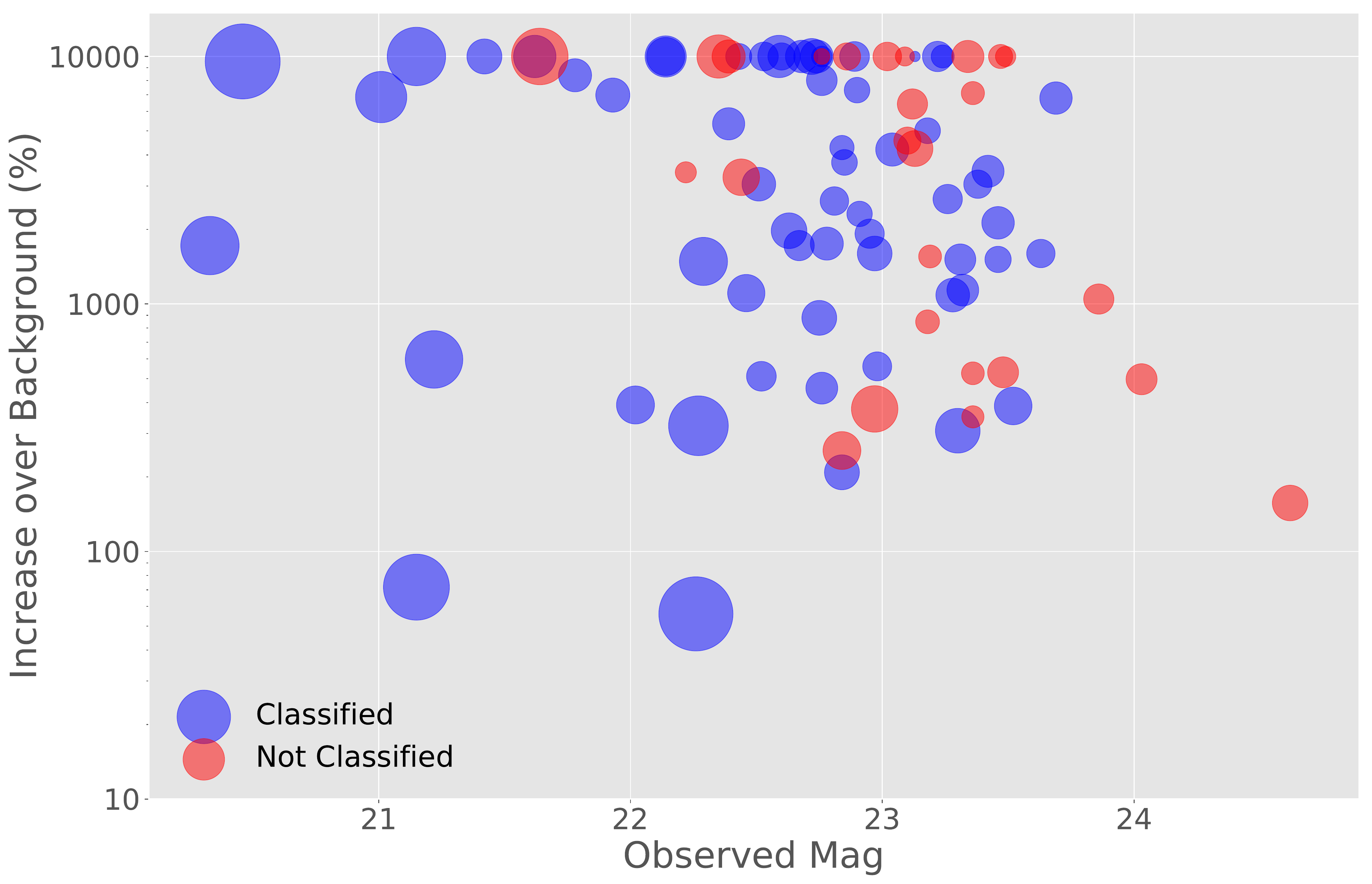}
	\caption{Observed $i$-band magnitude plotted against the percentage increase over background flux for each SN observed by our VLT program.  SNe that are $>10,000\%$ the brightness of their background are plotted at this value for purposes of clarity.  Successful classifications are shown in blue, while non-classification are plotted in red.  The size of each point is proportional to the SNR for the spectrum, with the method for computing this described in Section~\ref{sec:SpecSummary}.
	\label{fig:SNR_Classification_VLT}}
\end{figure}

We show in Figure~\ref{fig:SNR_Classification_VLT} the classified~vs.~unclassified spectra obtained at VLT, plotted as a function of apparent magnitude, percent increase over background, and SNR (indicated by the size of the points).  To compute the SNR for each spectrum we split the region $5000<\lambda({\textrm \AA})<9000$ into 200\,\AA\ sections, and determine the root-mean square (RMS) about the best-fit line in each section.  The SNR for the spectrum is then the average over all sections of the mean flux over RMS per section.  The linear fitting accounts for the fact that a SN spectrum has broad lines and therefore the simple RMS cannot be used as an indicator of uncertainty alone.  As expected, we find that the non-classified spectra tend to lie in the regime of low SNR and/or faint objects.  By design there are very few transients observed with this program that are not significantly above the background.  As a result our classification efficiency with VLT is high.

The number of spectroscopic classifications increased dramatically as the DES survey progressed. While the number of AAT nights increased modestly (from 10 to 12 to 16), the number of DES-SN spectra obtained from all other observatories over the first three seasons rose from 24 to 121 to 199.  This expansion of resources resulted in 24/75/152 spectroscopically classified \SNeIa\ in DES Y1/Y2/Y3, respectively.  We note that the efficiency of the survey pipeline itself also improved through the seasons; both the speed at which DES data were processed and the quality of artifact rejection increased dramatically from Y1 to Y3.  These improvements also contributed to the year-on-year increase in classified \SNeIa.

\begin{deluxetable}{lr}[h!]
\tablewidth{0pt}
\tabletypesize{\small}
\tablecaption{DES Classification Summary\label{tbl:classSum}}
\tablehead{
  \colhead{Type\hspace{2.0cm}}  &  
  \colhead{\hspace{2.0cm}Classifications} 
}
\startdata
SNIa     & 225 \\
SNIa?    & 26  \\
SNII     & 25  \\
SNII?    & 8   \\
SNIIn?   & 1   \\
SNIIb?   & 1   \\
SNIbc    & 5   \\
SNIc     & 4   \\
SLSN-I   & 11  \\
SLSN-I?  & 1   \\
TDE      & 1   \\
AGN      & 53 
\enddata
\end{deluxetable}

\section{Selection Function}
\label{sec:ssf}

One significant difficulty in an analysis of spectroscopically classified \SNeIa\ is the inherent complexity in determining how the observed sample relates to the broader set of photometric candidates: the spectroscopic selection function (SSF\footnote{Referred to as $E_{\textrm{spec}}$ by \citet{Kessler2018} and \citet{Brout2018-SYS}}).  In an idealized case, every SN would be classified.  This requires a telescope allocation large enough to target all transients, which is never realized (with the exception of very-low redshift surveys, such as ASAS-SN \citep{Shappee2014}), and thus requires some prioritization of transients.  Targeting is based on incomplete light curves.  Delays are introduced if many artifacts enter the photometry pipeline, or if data processing lags for any reason.  Telescope time is either classically scheduled or queue scheduled, neither of which guarantee a transient is observed at an appropriate epoch to maximize SNR and secure a classification.  Observing conditions -- clouds, seeing, moon phase -- all play a continuously varying role as well.  Available spectroscopic resources may be allocated in the wrong hemisphere for follow-up of a particular target.

In addition to accounting for the selection effects from {\tt DiffImg}, the inefficiencies of the spectroscopic sample must also be accounted for in the cosmology-parameter analysis.  Modeling all the inputs that ultimately shape our spectroscopic selection is not possible; this would have to incorporate computer failures that induced delays in data reduction as well as dust storms and instrument failures that caused prolonged shutdowns at certain observatories.  Rather than attempting to model the SSF from first principles, we determine an \emph{effective} SSF for \SNeIa\ in DES.  In Section~\ref{subsec:ssfDATA} we derive the SSF from our data.  In Section~\ref{subsec:ssfSIM} we use this SSF to create a DES-like simulation and compare the resulting \SNIa\ sample to the observed DES sample.  In Section~\ref{subsec:ssfMODEL} we examine a forward-modeling approach to deriving the SSF and how that differs from the data-driven model, and in Section~\ref{subsec:ssfBIAS} we show how these SSFs affect the distance-modulus bias correction for our spectroscopic \SNIa\ sample.

\subsection{SSF Derivation}
\label{subsec:ssfDATA}

The SSF is the fraction of \SNeIa\ identified as candidates by DES that are subsequently spectroscopically classified.  Thus, the denominator of the SSF should be the subset of single-season transients (SSTs; Section~\ref{sec:Cands}) detected by DES that were real \SNeIa.  Since we do not know the true classification of all SSTs from DES, we determine the likely number of \SNeIa\ based on the results of photometrically classifying our full three-year data sample (real-time decisions were based on classifications from rising light curves, which are less accurate than the full light curves used here).  We define the SSF as a function of the peak magnitude of the transient in the observer-frame $i$~band $(m_i)$, which is used for most spectroscopic follow-up decisions.  Thus,
\begin{equation}	
	\textrm{SSF}(m_i) = \frac{N_{\textrm{SpecIa}}(m_i)}{N_{\textrm{PhotIa}}(m_i)}.
	\label{eq:SSF}
\end{equation}

We select all 12,015 SSTs from the first three seasons of DES, and determine PBAYES$_{\textrm{Ia}}$ and FITPROB$_{\textrm{Ia}}$ using \PSNID\ (Section~\ref{sec:psnid}).  We use both the \SNIa\ light-curve models of \citet{Sako2011} as well as the set of core-collapse supernova (CCSN) templates adopted in that paper.  Motivated by the cut thresholds defined by \citet{Sako2011,Sako2018}, we select as likely \SNeIa\ all SSTs with PBAYES$_{\textrm{Ia}}\ge0.9$, FITPROB$_{\textrm{Ia}}\ge0.01$, and peak SNR$\ge5$ in at least two bands.  We note that our outlier rejection allows at most two filter-epochs of photometry to be removed from the fitting, provided they have $\Delta\chi^2 \ge 10$.  

For this subset of likely \SNeIa\ we determine the best-fit light-curve parameters using the Guy10\_UV2IR version of the SALT2 model, which was first used by \citet{Rodney2012} (for a description and a more advanced version of this model, see \citealt{Pierel2018}).  We note that this model is defined over a wider wavelength range than the B14-JLA model \citep{Betoule2014}, providing more complete coverage of the DES filters over a wide redshift range, and thereby allowing us to constrain the peak fit magnitude for most of our candidates.  We run our fits using the snlc\_fit module in the ``SuperNova ANAlysis" (\SNANA) software package \citep{Kessler2009b}, using no prior for the redshift for any \SNIa\ candidate.  We remove objects whose best-fit $x_1$ and color parameters are at the boundaries of our fitting range ($x_1=\pm4,c=\pm1)$, which typically signifies a non-convergent fit and a likely CCSN.  This leaves us with 2634 photometric \SNeIa.

\begin{figure}[!t]
	\centering
	\includegraphics[width=0.5\textwidth]{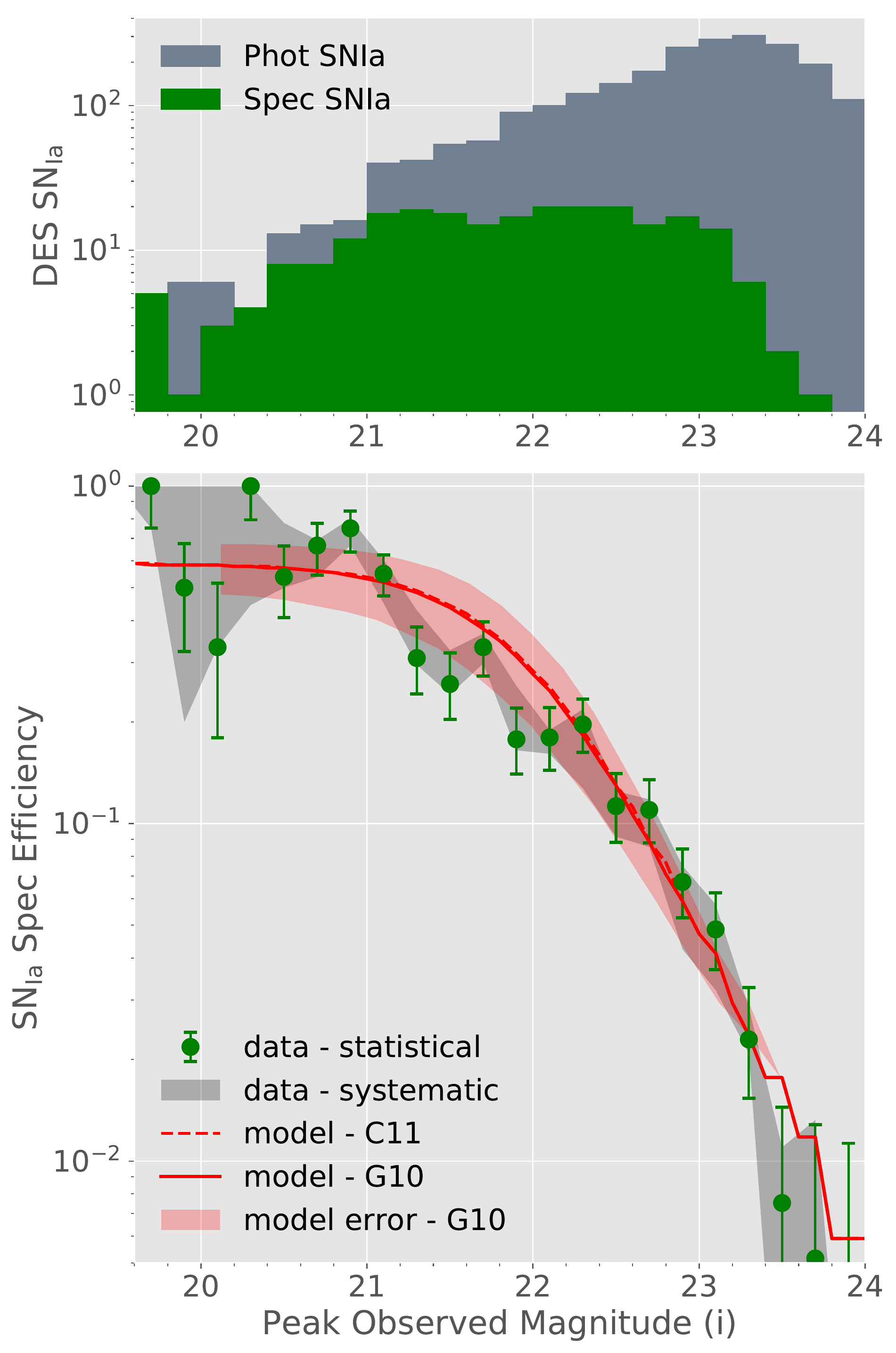}	\caption{\textbf{Top:} Histogram of the number of \SNeIa\ spectroscopically classified (251; green) and photometrically classified (2634; grey) in this paper, as a function of peak observed $i$-band magnitude. \textbf{Bottom:} Data-driven SSF (green), defined as a function of peak observed $i$-band magnitude, derived from the data shown in the top plot.  Error bars are statistical uncertainties in each bin derived from Eq.~(\ref{eq:error}), and the grey band is our estimated systematic uncertainty, described in Section~\ref{subsec:ssfDATA}.  In our analysis we set the efficiency to 1 at $i\leq20.4$ mag to prevent artifacts in our derived mu-bias, as described in the text.  The model-driven SSFs described in Section~\ref{subsec:ssfMODEL} are shown as solid and dotted lines depending on the assumed scatter model G10 or C11, respectively. We show the $1\sigma$ contour for the G10 derivation only, and note that the contours with C11 model are very similar.  The model-driven SSF has an arbitrary normalization, which we scale here to minimize the difference with respect to the data-driven SSF.
    \label{fig:SSF}}
\end{figure}

For the spectroscopic sample, we similarly fit each of the 251 spectroscopically classified \SNeIa\ with a Guy10\_UV2IR model and measure the best-fit, observer-frame $i$-band magnitude at peak.  We bin both our photometric and spectroscopic \SNeIa\ samples by peak $i$-band magnitude and divide to create our data-driven SSF, which we show in Figure~\ref{fig:SSF}.  The spectroscopic sample in its entirety is a subset of our photometric sample.

While photometry for all DES-SN candidates has been computed from the {\tt DiffImg} pipeline, we have also developed a photometric pipeline that forward models a variable transient flux on a temporally constant background.  This `Scene Modeling Photometry' \citep[\SMP;][]{Brout2018-PHOT} has been computed for the DES-SN3YR sample and is used in our cosmology analysis by \citet{Brout2018-SYS}, but is computed \emph{only} for these transients.  For the classifications and light-curve fits described above we have used \SMP\ for the spectroscopic sample, but find negligible differences in our results from using {\tt DiffImg} instead.  These differences are included in our systematic uncertainties, described in more detail below.

For magnitudes $m_i \leq 20.4$, we have set the efficiency equal to 1 (i.e., 100\% classification rate), despite the fact that there are \SNeIa\ in DES brighter than this limit which have not been confirmed.  We have made this choice because the SSF is ultimately used to determine how our selection creates a biased subset of \SNeIa.  The bright \SNeIa\ that went unconfirmed were not due to any lack of brightness, but rather operational issues (lack of telescope time or processing problems).  There are only a total of 34 photometric \SNeIa\ in this magnitude range---$1.3$\% of the total photometric sample.  We present our derived spectroscopic selection function in Table~\ref{tbl:SpecEff}.

\begin{deluxetable}{cccc}[h!]
\tablewidth{0pt}
\tabletypesize{\small}
\tablecaption{DES Spectroscopic Efficiency\label{tbl:SpecEff}}
\tablehead{
  \colhead{Peak Mag(i)}  &  
  \colhead{Efficiency} & 
  \colhead{$\sigma_\textrm{stat}$} &
  \colhead{$\sigma_\textrm{sys}$} 
}
\startdata
20.3  & 1.000 & +0.000/-0.205 & +0.000/-0.556 \\
20.5  & 0.538 & +0.127/-0.131 & +0.239/-0.038 \\
20.7  & 0.667 & +0.108/-0.121 & +0.026/-0.128 \\
20.9  & 0.750 & +0.094/-0.113 & +0.050/-0.083 \\
21.1  & 0.550 & +0.076/-0.078 & +0.055/-0.100 \\
21.3  & 0.309 & +0.073/-0.067 & +0.119/-0.014 \\
21.5  & 0.259 & +0.062/-0.056 & +0.067/-0.018 \\
21.7  & 0.333 & +0.063/-0.060 & +0.032/-0.035 \\
21.9  & 0.178 & +0.042/-0.038 & +0.078/-0.013 \\
22.1  & 0.180 & +0.040/-0.036 & +0.009/-0.019 \\
22.3  & 0.197 & +0.037/-0.034 & +0.021/-0.069 \\
22.5  & 0.113 & +0.028/-0.025 & +0.012/-0.021 \\
22.7  & 0.110 & +0.025/-0.022 & +0.008/-0.025 \\
22.9  & 0.067 & +0.017/-0.015 & +0.008/-0.025 \\
23.1  & 0.049 & +0.014/-0.012 & +0.009/-0.016 \\
23.3  & 0.023 & +0.010/-0.008 & +0.006/-0.002 \\
23.5  & 0.008 & +0.007/-0.004 & +0.004/-0.007 \\
23.7  & 0.005 & +0.008/-0.004 & +0.008/-0.001
\enddata
\tablecomments{For purposes of analysis, the efficiency is assumed to be 1.000 below magnitude 20.3.}
\end{deluxetable}

To determine the statistical uncertainty in each bin, we follow the method adopted by \citet{Frohmaier2017} for supernova rate calculations, which draws upon \citet{Paterno2004}.  In each bin the number of spectroscopic classifications $k$ out of $n$ detected \SNeIa\ is a binomially distributed variable, with the Bayesian posterior probability distribution for the rate $\epsilon$ defined as
\begin{equation}
\label{eq:error}
p(\epsilon|k,n) = \frac{\Gamma(n+2)}{\Gamma(k+1)\Gamma(n-k+1)}\epsilon^{k}(1-\epsilon)^{n-k}.
\end{equation}
We plot as our error bars in Figure~\ref{fig:SSF} the 1$\sigma$ uncertainties, i.e., the bounds containing $68.3$\% of the probability.

In deriving the selection function there were several choices made that could have been done differently.  We could have fit the JLA-B14 model to determine the peak magnitude rather than Guy\_UV2IR; we could have used the observed rather than fit peak magnitude; we also could have used higher or lower \PSNID\ cut thresholds for defining a likely \SNIa.  We have re-computed the SSF using each of these variations to our analysis, including in the PBAYES$_{\textrm{Ia}}$ (0.5, 0.99) and FITPROB$_{\textrm{Ia}}$ (0.001, 0.1) parameters.  We define as our systematic uncertainty the maximum variation in our derived SSF in each magnitude bin, which we plot in Figure~\ref{fig:SSF}.  We note that the systematic uncertainty is comparable to or smaller than the statistical uncertainty over nearly the entire magnitude range to which we are sensitive.

\subsection{Data Comparison to Simulations}
\label{subsec:ssfSIM}

The SSF derived above comes directly from the observed data, and is not dependent on simulations or any assumptions about \SNIa\ rates or intrinsic parameter distributions.  However, it is useful as a consistency check to determine whether our derived SSF produces a spectroscopic \SNIa\ sample similar in redshift distribution and \SNIa\ light-curve parameters to that observed when input into a DES-like simulation.

As part of the first DES-SN cosmology analysis \citep{DES-SN2018, Brout2018-SYS} -- which uses only spectroscopically confirmed \SNeIa\ from the first three years of the survey, described in this paper -- we have created a suite of simulations \citep{Kessler2018} that produces a highly accurate DES-like survey.  We refer the reader to Figure~1 of \citet{Kessler2018} for an overview of the \SNANA\ simulation, and list below some of the more important assumed parameters:

\begin{itemize}
  \setlength\itemsep{0em}
  \item SALT2 model: \citet{Betoule2014} (JLA--B14\_LAMOPEN)
  \item \SNIa\ instrinsic scatter model: \citet{Guy2010}
  \item \SNIa\ rates: \citet{Perrett2012}
  \item SALT2 parameter distribution: \citet{Scolnic2016} (Table 1, high-$z$, G10 row)  
  \item MW dust maps: \citet{Schlafly2011}
  \item Extinction law: \citet{Fitzpatrick1999}
  \item Nuisance parameters: $\alpha=0.15, \beta=3.1$
  \item Cosmology: Flat $\Lambda$CDM, $\Omega_M=0.3$, H$_0=70$~km~s$^{-1}$~Mpc$^{-1}$
\end{itemize}

\begin{figure*}[htp]
	\centering
	\includegraphics[width=1.0\textwidth]{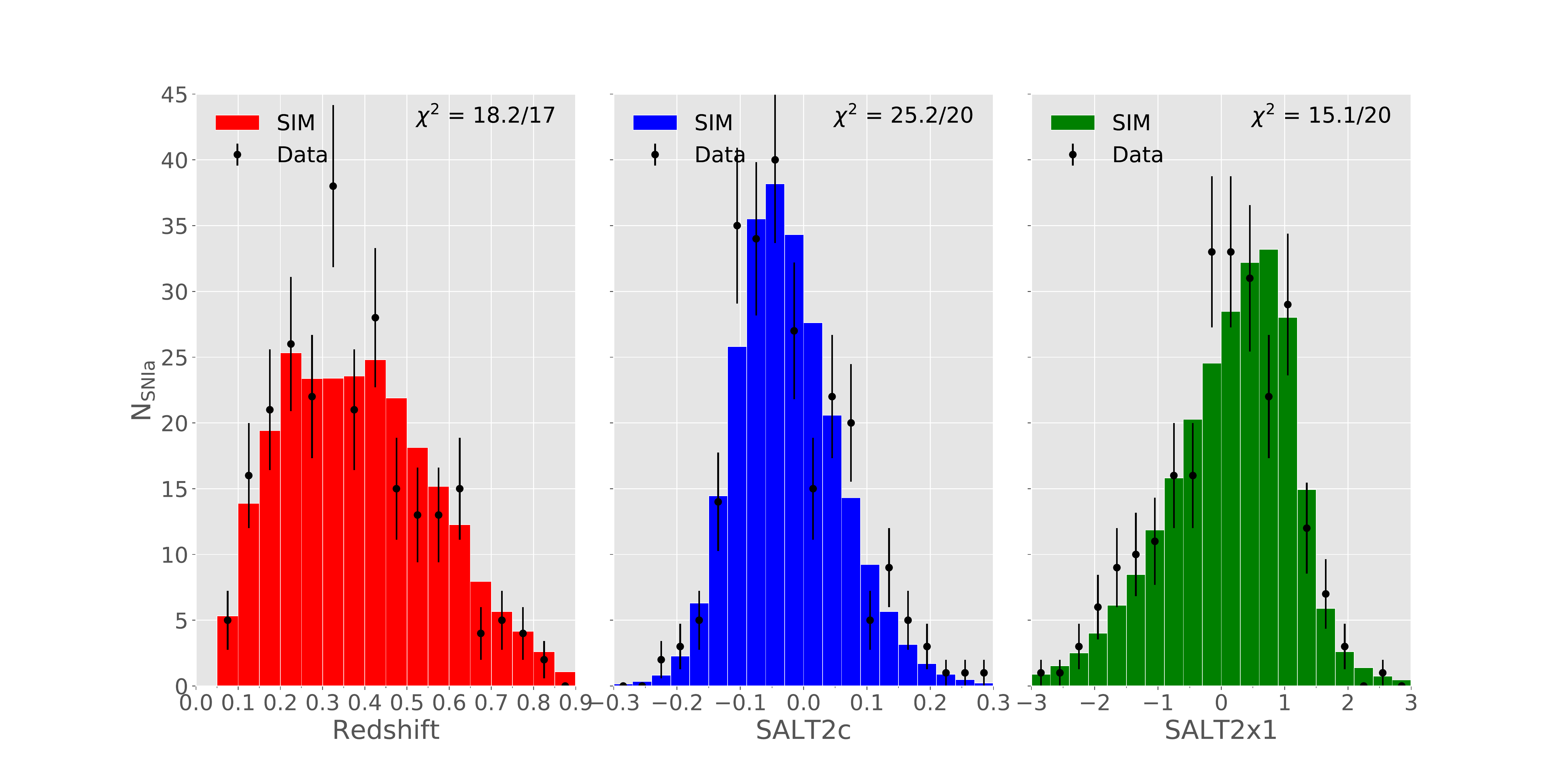}
	\caption{Distributions of redshift, SALT2 $x_1$, and SALT2 $c$ for both our spectroscopic sample and a DES-like simulation using the SSF we derived from the data (and which therefore should be representative of the spectroscopic sample).  The simulation is normalized to the number of points in the data histogram, and uncertainties on the data are $\sqrt{N}$ statistics.  The goodness-of-fit for each histogram is shown as the $\chi^2$ on each plot.
	\label{fig:CompHist}}
\end{figure*}

The SALT2 parameters are computed using the JLA--B14\_LAMOPEN model{\footnote{This is an extension in wavelength coverage of the JLA--B14 model, which would otherwise only use DES-SN $iz$ photometry in light-curve fits of low-redshift ($z\le0.11$) SNe.}}, and is performed on both \SMP\ photometry and simulations.  We show the binned redshift, $x_1$, and color distribution for the spectroscopic sample, compared to the simulated sample, in Figure~\ref{fig:CompHist}.  We note that the SSF used in the simulation shown here is the one we have derived from the data; \citet{Kessler2018} use a forward-modelling SSF, which we describe in Section~\ref{subsec:ssfMODEL}.  The simulation histograms have been normalized to the same number of \SNeIa\ as the spectroscopic sample.  We use $\sqrt{N}$ errors for the observed data, and list the $\chi^2$/DoF for the fit between the binned simulations and data.  

There is good agreement between our simulation and the data.  The $\chi^2$/DoF for the redshift, $x_1$, and color distributions indicate agreement between the data and simulations.  This does not simply represent a judicious choice of binning; for example, shifting the starting point of our $\Delta c=0.03$ bins by 0.01 in either direction leads to $\chi^2$/DoF$<1$.  We have also run a two-sided Kolmogorov-Smirnov (KS) test using the distributions of these parameters, which for small p-values would be able to rule out the hypothesis that the samples are drawn from the same parent distribution.  We find p-values of 0.11, 0.38, and 0.16 for $z$, $x_1$, and $c$, respectively, which are consistent with coming from the same distribution.

\begin{figure}[htp]
	\centering
	\includegraphics[width=0.5\textwidth]{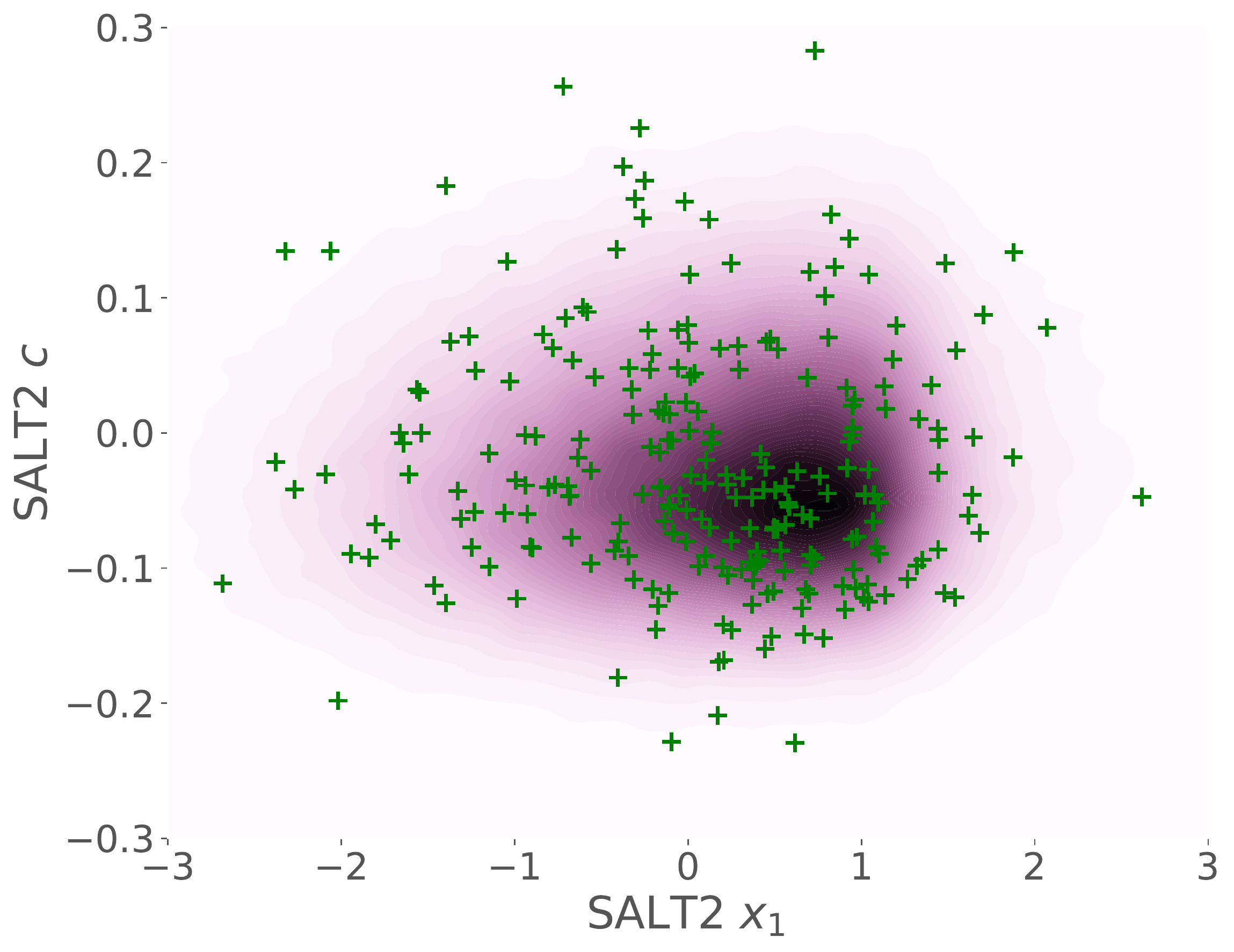}
	\caption{Joint distribution of SALT2 $x_1$ and $c$ for a DES-like simulation that uses the data-driven SSF.  The contours are derived from a kernal density estimator, where darker colors represent higher population density.  Measured parameters from our spectroscopically confirmed \SNeIa\ are plotted in green. 
	\label{fig:x1c2D}}
\end{figure}

\begin{figure}[htp]
	\centering
	\includegraphics[width=0.5\textwidth]{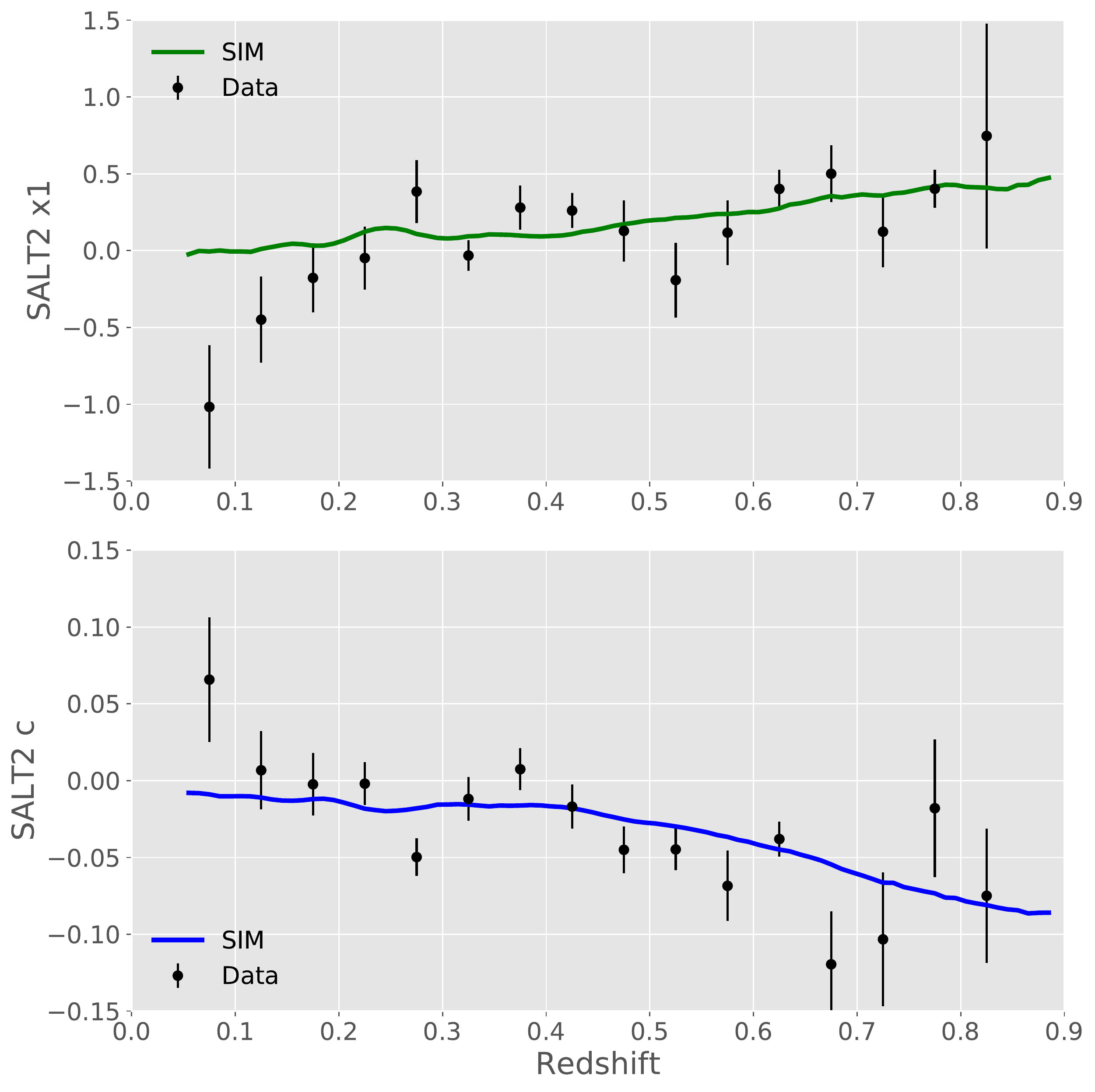}
	\caption{Redshift evolution of SALT2 $x_1$ and $c$ for a DES-like simulation that uses the data-driven SSF.  The lines are rolling averages of the simulated parameters, while weighted mean and the standard deviation on the mean are shown for the data, binned by $\Delta z=0.05$.
	\label{fig:zDistComp}}
\end{figure}

In Figure~\ref{fig:x1c2D} we plot the two-dimensional distribution of derived SALT2 $x_1$ and $c$ from our spectroscopic sample, and compare it to a contour plot derived from a kernel density estimator (KDE) of our simulation.  In Figure~\ref{fig:zDistComp} we show the simulated and measured evolution of SALT2 parameters over the redshift range of our observations.  Both plots demonstrate that our data generally match the trends expected from our simulation, with our $x_1/c$ measurements clustered around the peak of the KDE and an increasing (decreasing) trend with redshift of SALT2 $x_1$ ($c$).  There are some statistically significantly differences: an overabundance of observed \SNeIa\ in low-probability areas of $x_1/c$ space, and a trend for lower-than-expected $x_1$ values at low~$z$.  It is difficult to state whether these differences are signs of unaccounted bias in the spectroscopic sample selection, or rather the intrinsic parameter model for \SNeIa\ in DES differs from that determined from SDSS+PS1+SNLS data \citep{Scolnic2016}.  

Our SSF in Eq.~(\ref{eq:SSF}) is clearly a simplified description of our strategy.  Supernova follow-up decisions and classification efficiency (both of which are combined into the SSF) were functions of multiple variables beyond peak magnitude, including host-galaxy mass, local surface brightness, and observed color.  We explored adding these variables to our SSF definition, but as shown in this Section, using just a single variable provides an effective description of the full DES-SN3YR spectroscopic sample.

\subsection{Simulation-Based Derivation}
\label{subsec:ssfMODEL}

In Section~\ref{subsec:ssfDATA} we inferred the effective selection function directly from the candidates detected in the data.  This requires making assumptions about our ability to accurately photometrically classify SN candidates from our full set of SSTs.  While we have shown in Section~\ref{subsec:ssfSIM} that this method succeeds in reproducing the majority of the trends seen in our spectroscopic \SNIa\ sample, an alternative approach to the SSF derivation would be a useful check.

Using simulations without any spectroscopic selection function, one can infer the expected magnitude distribution of real \SNeIa\ in our photometric data.  This quantity can then be used as the `truth' in the denominator of the SSF derivation.  We call this approach the model-driven selection function to differentiate from the data-driven selection function of Section~\ref{subsec:ssfDATA}.

This method has the benefit of being insensitive to the various uncertainties that could potentially bias our photometric classification, led by contamination from CCSNe.  It is also highly dependent on the input model.  This includes the distribution of \SNIa\ SALT2 parameters ($x_1, c$), their rate as a function of redshift, and the intrinsic scatter model that determines the irreducible variation in the \SNIa\ Hubble diagram ($\sigma_{\textrm{intrinsic}}$).  We note that the values assumed for the \SNIa\ population by \citet{Kessler2018} are commonly used, and that extensive testing on artificial SNe inserted into images by \citet{Kessler2018} has shown that the simulations produce DES-like data. 

In Figure~\ref{fig:SSF} we show the selection functions determined using this method, with the intrinsic scatter on the simulated SNe assumed from either the intrinsic scatter models of \citet{Guy2010} (G10) or \citet{Chotard2011} (C11).  The assumption of either scatter model results in minimal differences in the derived SSF, so for clarity we only overplot the $1\sigma$ errors on the model assuming G10.  The model-driven SSF efficiency $\epsilon$ is determined from a sigmoid fit to the binned data,
\begin{equation}
\label{eq:sigmoid}
\epsilon(i_{\textrm{peak}}) = \frac{s_0}{1+\exp(s_{1}\times i_{\textrm{peak}} - s_2)},
\end{equation}
where $s_0$, $s_1$, and $s_2$ are free parameters determined with {\tt emcee} \citep{Foreman-Mackey2013}.  For the fitting, we model data uncertainties using a Poisson distribution.  We perform the sigmoid fit on the data-simulation ratio and then use the sigmoid amplitude, $s_0$, as normalization parameter. This allows the selection function to asymptote to a constant value for bright transients, go to zero for sufficiently faint transients, and transition smoothly between the two.

The two derivations of the selection function largely agree.  Using the statistical uncertainties from the data-driven model and the $1\sigma$ contour from the model-based derivation, the $\chi^2$/DoF $=0.7$.  Visible differences between the two selection functions may be due to the assumption of a smooth sigmoid function; while classification efficiency does monotonically increase for brighter transients on a given telescope, the strategy of having different telescopes for different targets results in a non-smoothly varying process.  We run a DES-like simulation assuming this selection function, and again compare the resulting $z$, $x_1$, and $c$ distributions to the data using the two-sample KS test.  We find the probabilities that these distributions are consistent with one another to be 0.02, 0.41, and 0.18, respectively.  While the $x_1$ and $c$ distributions are consistent, there is some evidence for a difference in the redshift distribution, primarily driven by an underabundance of \SNeIa\ in the simulation compared to data at low~$z$.  This underabundance, which peaks at $z\approx0.25$, is similar in nature but slightly stronger than the underabundance at mid-low redshifts from the data-driven model, which peaks at $z\approx0.35$.

We are primarily interested in the two derivations insofar as they inform our systematic uncertainty on our cosmology analysis of the spectroscopically classified sample; this is evaluated in the following section. 

\subsection{Mu-Bias}
\label{subsec:ssfBIAS}

The cosmological parameter analysis and measurement of systematic uncertainties for the DES-SN3YR sample in \citet{Brout2018-SYS} uses the model-driven SSF as a baseline.  The utility of this approach is clear, as the SSF is derived from the same suite of simulations defined by \citet{Kessler2018} and used to quantify the uncertainties in the analysis.  It would also be difficult to do a rigorous evaluation of systematic uncertainties owing to the photometrically-classified sample used in the data-driven approach.  However, the benefit of the data-driven approach is that, being free of the assumptions that go into defining the model of the observed \SNeIa\ population, it allows for a cross-check of the SSF that cannot be computed from the simulations alone.  \citet{Brout2018-SYS} use a $1\sigma$ statistical fluctuation on the model-driven SSF and propagate it as a systematic uncertainty, finding the SSF to be only the ninth largest source of uncertainty in the equation-of-state parameter $w$.

\begin{figure}[htp]
	\centering
	\includegraphics[width=0.5\textwidth]{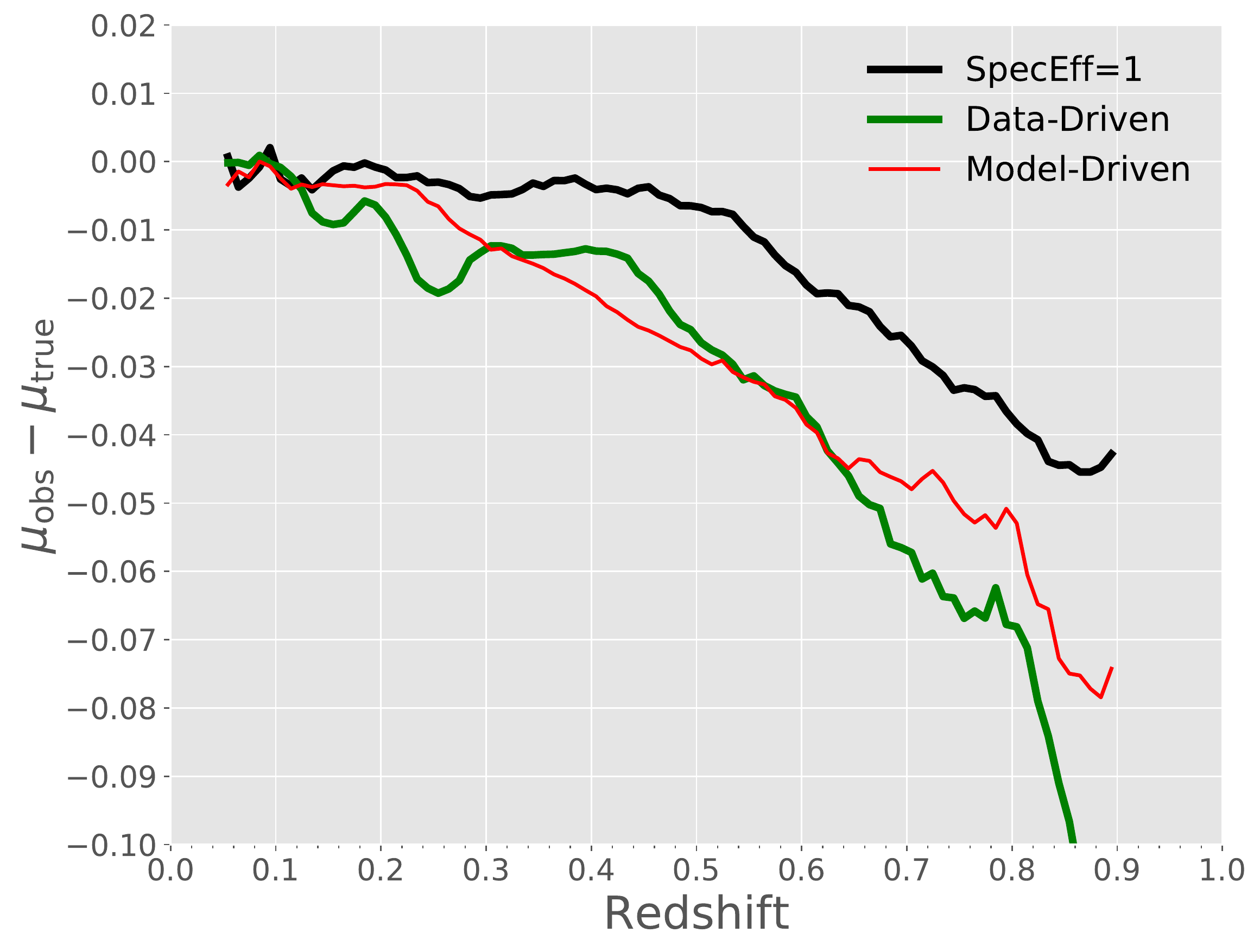}
	\caption{Redshift dependent bias derived from Spectroscopic Selection Function derivations.  Black solid line shows the mu-bias derived with perfect spectroscopic efficiency, which reveals the limitations of the survey (and perfect photometric classification).  The green line is the mu-bias derived from the data-driven SSF, while the red line shows the model-driven SSF.  All assume an intrinsic G10 scatter model for consistency.
	\label{fig:mubias}}
\end{figure}

We demonstrate the effects the SSF imparts on our observed data in Figure~\ref{fig:mubias}, where we show the simulated redshift-dependent bias in distance modulus for various assumptions about our selection effects.  The bias ($\Delta\mu$) is computed as the difference between the distance modulus derived from fitting light curves ($\mu_{\textrm obs}$) and the true distance modulus ($\mu_{\textrm true}$).  Each is derived from an identical simulation where the only difference is the assumed SSF.  The black solid line shows the mu-bias that would be expected from a perfect spectroscopic selection; the depth of the photometry from DES means there would be nearly no bias for a perfectly selected \SNIa\ sample out to $z\approx0.45$, smoothly dropping off thereafter to a bias of $\Delta\mu=0.045$ at $z\approx0.85$.  Refer to Figure~\ref{fig:zHist} for the redshift distribution of the DES-SN3YR subsample.

The data-driven and model-driven functions (shown in Figure~\ref{fig:mubias} in green and red, respectively) agree on average, differing by $>0.01$ mag only at the lowest and highest redshifts.  The differences can be attributed to the different shapes of the selection functions (Figure~\ref{fig:SSF}).  The steps and plateaus seen in the data-driven model are due to the binned nature of the data-driven model, while the functional form of the model-driven SSF leads to a simpler redshift evolution: flat at low $z$, and an effectively linear decline thereafter.  The difference in mu-bias between the black line and either of the two other lines isolates the effect on the mu-bias due to the spectroscopic selection, distinct from the pipeline detection efficiency.  This demonstrates that the mu-bias due explicitly to the program described in this paper lies between $0.01$ to $0.03$ mag over a wide range of redshift.

\section{Conclusion}
\label{sec:conc}

In this paper we have presented the survey operations and spectroscopic follow-up observations for the first three years of the Dark Energy Survey - Supernova Program.

We presented a detailed overview of the DES-SN observing strategy: exposure times, depths, image quality, field locations, and cadence.  On average, a DES season was $\sim5.5$ months long, with each of our 10 fields observed in $griz$ every 7.4\,d.  The median depth was $\sim23.5$ mag in the eight shallow fields and $\sim24.6$ mag in the two deep fields, with wider variance on a per-image basis for the bluer bands.  As the best seeing typically (but not always) went to DES-wide, the median observed FWHM in $griz$ for our program was $1.41\arcsec/1.29\arcsec/1.17\arcsec/1.09\arcsec$.

We described results from our data processing pipeline, the details of which are mostly contained in other papers \citep{Kessler2015,Goldstein2015,Gupta2016,Morganson2018}.  DES-SN recorded 1.21 million real single-epoch, single-filter detections---nearly 400 per image.  From these detections approximately 46,000 supernova candidates were identified, which we subsequently narrowed down to 12,015 viable single-season transients.  On average we discovered 24 likely supernovae per day.

We have given an overview of our live-supernova spectroscopy follow-up, consisting of Magnitude-Limited, Faint Host, Representative, and Non-Ia programs.  Observations were made on an assortment of telescopes.  The largest fraction by far comes from multi-object spectroscopy with the `OzDES' on the AAT, with large allocations on VLT and Magellan and significant follow-up from Gemini, Keck, GTC, SALT, and MMT. In total we collected 1352 spectra -- 1009 of which were from AAT -- resulting in 251 classifications of type Ia supernovae and 56 non-Ia SNe.  Our spectroscopically classified \SNIa\ sample spans a redshift range of $0.017 < z < 0.85$.  Analyses of the SNe collected in each of these individual samples will be presented in other DES-SN papers.  The host-galaxy follow-up of DES transients to obtain spectroscopic redshifts for a photometric sample -- which is the primary goal of OzDES -- will also be described in a future paper.

We derived the effective spectroscopic selection function (SSF) from our large, diverse follow-up program that resulted in our classified sample of DES \SNeIa.  One method is data-driven, relying on photometric classification to determine the fraction of real \SNIa\ in our data, while the model-driven method relies on simulations of our survey.  The two methods produce distinctly different shapes for our effective SSF, underscoring the different assumptions of the two models and the limitations of accurately characterizing the SSF.  We show the resulting redshift-dependent bias the SSF imparts upon the measured distance modulus for each method, and remark on how this is a subdominant systematic error on the resulting cosmology analysis, as shown in the companion paper \citet{DES-SN2018}.

\acknowledgments
This paper has gone through internal review by the DES collaboration.

Based in part on data acquired through the Australian Astronomical Observatory under program ATAC A/2013B/12. We acknowledge the traditional owners of the land on which the AAT [or UKST] stands, the Gamilaraay people, and pay our respects to elders past and present.  Based on observations obtained at the Gemini Observatory, which is operated by the Association of Universities for Research in Astronomy, Inc., under a cooperative agreement with the NSF on behalf of the Gemini partnership: the National Science Foundation (United States), the National Research Council (Canada), CONICYT (Chile), Ministerio de Ciencia, Tecnolog\'{i}a e Innovaci\'{o}n Productiva (Argentina), and Minist\'{e}rio da Ci\^{e}ncia, Tecnologia e Inova\c{c}\~{a}o (Brazil).  Observations with Gemini were obtained under NOAO programs 2013A-0373 and 2015B-0197, corresponding to GN-2013B-Q-55, GS-2013B-Q-45, GS-2015B-Q-7, GN-2015B-Q-10, as well as GS-2015B-Q-8 under a Chilean program.  Based on observations made with the Gran Telescopio Canarias (GTC), installed at the Spanish Observatorio del Roque de los Muchachos of the Instituto de Astrof\'{i}sica de Canarias, in the island of La Palma.  Observations with GTC were made under programs GTC77-13B, GTC70-14B, and GTC101-15B.  Some of the data presented herein were obtained at the W. M. Keck Observatory, which is operated as a scientific partnership among the California Institute of Technology, the University of California and the National Aeronautics and Space Administration. The Observatory was made possible by the generous financial support of the W. M. Keck Foundation.  Observations with Keck were made under programs U063-2013B, U021-2014B, U048-2015B, and U038-2016A.  The authors wish to recognize and acknowledge the very significant cultural role and reverence that the summit of Maunakea has always had within the indigenous Hawaiian community.  We are most fortunate to have the opportunity to conduct observations from this mountain.  This paper includes data gathered with the 6.5 meter Magellan Telescopes located at Las Campanas Observatory, Chile, partially through program CN2015B-89.  Observations reported here were obtained at the MMT Observatory, a joint facility of the Smithsonian Institution and the University of Arizona, under programs 2014c-SAO-4, 2015a-SAO-12, 2015c-SAO-21.  Some of the observations reported in this paper were obtained with the Southern African Large Telescope (SALT) under programs 2013-1-RSA\_OTH-023, 2013-2-RSA\_OTH-018, 2014-1-RSA\_OTH-016, 2014-2-SCI-070, 2015-1-SCI-063, and 2015-2-SCI-061.  Based on observations collected at the European Southern Observatory under ESO programmes 093.A-0749(A), 094.A-0310(B), 095.A-0316(A), 096.A-0536(A), 095.D-0797(A).  Based on observations obtained at the Southern Astrophysical Research (SOAR) telescope, which is a joint project of the Minist\'{e}rio da Ci\^{e}ncia, Tecnologia, Inova\c{c}\~{o}es e Comunica\c{c}\~{o}es (MCTIC) do Brasil, the U.S. National Optical Astronomy Observatory (NOAO), the University of North Carolina at Chapel Hill (UNC), and Michigan State University (MSU).  SOAR observations obtained under program 2014B-0205.  Research at Lick Observatory is partially supported by a generous gift from Google.

The Penn group was supported by DOE grant DE-FOA-0001358 and NSF grant AST-1517742, and the Southampton group acknowledges support from EU-FP7/ERC grant [615929].  This research used resources of the National Energy Research Scientific Computing Center (NERSC), a DOE Office of Science User Facility supported by the Office of Science of the U.S. Department of Energy under Contract No. DE-AC02-05CH11231.  A.V.F.'s group at U.C. Berkeley is grateful for financial assistance from NSF grant AST-1211916, the Christopher R. Redlich Fund, Gary and Cynthia Bengier, the TABASGO Foundation, and the Miller Institute for Basic Research in Science.  The UCSC team is supported in part by NASA grants 14-WPS14-0048, NNG16PJ34G, NNG17PX03C, NSF grants AST-1518052 and AST-1815935, the Gordon \& Betty Moore Foundation, the Heising-Simons Foundation, and by fellowships from the Alfred P.\ Sloan Foundation and the David and Lucile Packard Foundation to R.J.F.  

Funding for the DES Projects has been provided by the U.S. Department of Energy, the U.S. National Science Foundation, the Ministry of Science and Education of Spain, the Science and Technology Facilities Council of the United Kingdom, the Higher Education Funding Council for England, the National Center for Supercomputing Applications at the University of Illinois at Urbana-Champaign, the Kavli Institute of Cosmological Physics at the University of Chicago, the Center for Cosmology and Astro-Particle Physics at the Ohio State University, the Mitchell Institute for Fundamental Physics and Astronomy at Texas A\&M University, Financiadora de Estudos e Projetos, Funda{\c c}{\~a}o Carlos Chagas Filho de Amparo {\`a} Pesquisa do Estado do Rio de Janeiro, Conselho Nacional de Desenvolvimento Cient{\'i}fico e Tecnol{\'o}gico and the Minist{\'e}rio da Ci{\^e}ncia, Tecnologia e Inova{\c c}{\~a}o, the Deutsche Forschungsgemeinschaft, and the Collaborating Institutions in the Dark Energy Survey. 

The Collaborating Institutions are Argonne National Laboratory, the University of California at Santa Cruz, the University of Cambridge, Centro de Investigaciones Energ{\'e}ticas, Medioambientales y Tecnol{\'o}gicas-Madrid, the University of Chicago, University College London, the DES-Brazil Consortium, the University of Edinburgh, the Eidgen{\"o}ssische Technische Hochschule (ETH) Z{\"u}rich, Fermi National Accelerator Laboratory, the University of Illinois at Urbana-Champaign, the Institut de Ci{\`e}ncies de l'Espai (IEEC/CSIC), the Institut de F{\'i}sica d'Altes Energies, Lawrence Berkeley National Laboratory, the Ludwig-Maximilians Universit{\"a}t M{\"u}nchen and the associated Excellence Cluster Universe, the University of Michigan, the National Optical Astronomy Observatory, the University of Nottingham, The Ohio State University, the University of Pennsylvania, the University of Portsmouth, SLAC National Accelerator Laboratory, Stanford University, the University of Sussex, Texas A\&M University, and the OzDES Membership Consortium.

Based in part on observations at Cerro Tololo Inter-American Observatory, National Optical Astronomy Observatory, which is operated by the Association of Universities for Research in Astronomy (AURA) under a cooperative agreement with the National Science Foundation.  

The DES data management system is supported by the National Science Foundation under Grant Numbers AST-1138766 and AST-1536171.  The DES participants from Spanish institutions are partially supported by MINECO under grants AYA2015-71825, ESP2015-66861, FPA2015-68048, SEV-2016-0588, SEV-2016-0597, and MDM-2015-0509, some of which include ERDF funds from the European Union. IFAE is partially funded by the CERCA program of the Generalitat de Catalunya.  Research leading to these results has received funding from the European Research Council under the European Union's Seventh Framework Program (FP7/2007-2013) including ERC grant agreements 240672, 291329, and 306478.  We  acknowledge support from the Australian Research Council Centre of Excellence for All-sky Astrophysics (CAASTRO), through project number CE110001020, and the Brazilian Instituto Nacional de Ci\^encia e Tecnologia (INCT) e-Universe (CNPq grant 465376/2014-2).

This manuscript has been authored by Fermi Research Alliance, LLC under Contract No. DE-AC02-07CH11359 with the U.S. Department of Energy, Office of Science, Office of High Energy Physics. The United States Government retains and the publisher, by accepting the article for publication, acknowledges that the United States Government retains a non-exclusive, paid-up, irrevocable, world-wide license to publish or reproduce the published form of this manuscript, or allow others to do so, for United States Government purposes.


\end{document}